\documentclass[aps,prl,floatfix,twocolumn,superscriptaddress,reprint]{revtex4-2}
\usepackage{amsmath}
\usepackage{amssymb}
\usepackage{eurosym}
\usepackage{amsbsy}
\usepackage{subfigure}
\usepackage{comment}
\usepackage{color}
\usepackage{mathrsfs}
\usepackage{amsfonts}
\usepackage{xspace}
\usepackage{latexsym,epsfig,graphicx}
\usepackage{epstopdf}
\usepackage{tabularx}
\usepackage{longtable}
\usepackage{bm, amsfonts, braket}
\usepackage{amsthm}
\usepackage{titlesec}
\newtheorem{thm}{Theorem}
\usepackage[normalem]{ulem}
\usepackage[version=4]{mhchem}
\usepackage{multirow}
\usepackage[mathlines]{lineno}
\usepackage{dcolumn}
\usepackage[colorlinks=true, pdfstartview=FitV, linkcolor=blue, citecolor=blue, urlcolor=blue]{hyperref}
\usepackage{bm}

\newcommand\wrapped[1]%
  {%
   \begin{array}{@{}l@{}}#1\end{array}%
  }
\usepackage{xcolor}

\begin{document}
\preprint{APS/123-QED}

\title{Braiding topology of symmetry-protected degeneracy points in non-Hermitian systems}

\author{Jia-Zheng Li}
\affiliation{Key Laboratory of Artificial Micro- and Nano-structures of Ministry of Education and School of Physics and Technology, Wuhan University, Wuhan 430072, China}
\author{Kai Bai}
\affiliation{Key Laboratory of Artificial Micro- and Nano-structures of Ministry of Education and School of Physics and Technology, Wuhan University, Wuhan 430072, China}
\author{Cheng Guo}
\affiliation{Department of Applied Physics, Stanford University, Stanford, California 94305, USA}
\author{Tian-Rui Liu}
\affiliation{Key Laboratory of Artificial Micro- and Nano-structures of Ministry of Education and School of Physics and Technology, Wuhan University, Wuhan 430072, China}
\author{Liang Fang}
\affiliation{Key Laboratory of Artificial Micro- and Nano-structures of Ministry of Education and School of Physics and Technology, Wuhan University, Wuhan 430072, China}
\author{Duanduan Wan}
\email{ddwan@whu.edu.cn}
\affiliation{Key Laboratory of Artificial Micro- and Nano-structures of Ministry of Education and School of Physics and Technology, Wuhan University, Wuhan 430072, China}
\author{Meng Xiao}
\email{phmxiao@whu.edu.cn}
\affiliation{Key Laboratory of Artificial Micro- and Nano-structures of Ministry of Education and School of Physics and Technology, Wuhan University, Wuhan 430072, China}
\affiliation{Wuhan Institute of Quantum Technology, Wuhan 430206, China}

\begin{abstract}
Degeneracy points in non-Hermitian systems are of great interest. While a homotopic framework exists for understanding their behavior in the absence of symmetry, it does not apply to symmetry-protected degeneracy points with reduced codimension. In this work, utilizing algebraic topology, we provide a systematic classification of these symmetry-protected degenerate points and investigate the braid conservation rule followed by them. Using a model Hamiltonian and circuit simulation, we discover that, contrary to simple annihilation, pairwise created symmetry-protected degeneracy points merge into a higher order degeneracy point, which goes beyond the abelian picture. Our findings empower researchers across diverse fields to uncover new phenomena and applications harnessing symmetry-protected non-Hermitian degeneracy points.
\end{abstract}

\maketitle

{\color{blue}\textit{Introduction.—}}
Band degeneracies have played a significant role in topological band theory of Hermitian systems, with their topology classified through homotopy theory \cite{NonAbelianBandTopology2019wu,TopologicalNodalLine2015fang,RobustDoublyCharged2017bzdusek}. Well-known instances, such as Weyl points, Dirac points and nodal lines~\cite{TopologicalNodalLine2016fang,WeylDiracSemimetals2018armitagea,WeylDiracHighfold2021hasan,TopologicalMaterialsWeyl2017yan,TypeIIWeylSemimetals2015soluyanov}, lead to a plethora of exotic physics~\cite{AccidentalDegeneracyPhotonic2016xu,SecondorderTopologyMultidimensional2019zhang,ObservationChiralZero2019jia,QuantumNonlinearHall2015sodemann,MomentumspaceSignaturesBerry2021unzelmann,NonAbelianReciprocalBraiding2020bouhon,ExperimentalObservationNonAbelian2021guo}. In recent years, the study of non-Hermitian systems has gained momentum~\cite{SymmetryTopologyNonHermitian2019kawabata,ClassificationExceptionalPoints2019kawabata,ExceptionalTopologyNonHermitian2021bergholtz,NonHermitianTopologyExceptionalpoint2022ding,ExceptionalPointsOptics2019miri,TopologicalComplexenergyBraiding2021wang,SymmetryprotectedExceptionalNodal2022sayyad}. In non-Hermitian settings, degeneracies can possess complex values and encompass more exotic singularities, such as defective degeneracies known as exceptional points~\cite{EnhancedSensitivityHigherorder2017hodaei,NonlinearExceptionalPoints2023bai,ObservationChiralState2022nasari}, as well as unique non-defective degeneracy points that do not have counterparts in the Hermitian regime~\cite{SymmetryprotectedExceptionalNodal2022sayyad,ProtectionAllNondefective2022sayyad,NonHermitianPhysics2020ashida}.
Recent investigations have revealed that non-Hermitian degeneracies without any symmetry can be classified by the braid group $\mathcal{B}_n$~\cite{HomotopyCharacterizationNonHermitian2020wojcik,HomotopicalCharacterizationNonHermitian2021li}, which goes beyond the topological classification based on line or point gaps~\cite{SymmetryTopologyNonHermitian2019kawabata,ClassificationExceptionalPoints2019kawabata}. 
Since $\mathcal{B}_n$ is a non-abelian group for $n\ge3$ where $n$ is the dimension of a Hamiltonian, the evolution of non-Hermitian degeneracies follows a non-abelian conservation rule (NACR)~\cite{ExceptionalNonAbelianTopology2023guo,KnotTopologyExceptional2022hu,BraidprotectedTopologicalBand2023konig}. Braided structures in Hermitian systems have led to many exotic phenomena~\cite{TyingKnotsLight2013kedia,KnottingFractionalorderKnots2019pisanty,KnottedThreadsDarkness2004leach,NonAbelianBandTopology2019wu,ImagingNodalKnots2020lee,DoubleHelixNodal2017sun,QuantumFieldTheory1989witten,NonAbelianReciprocalBraiding2020bouhon,FloquetMultigapTopology2022slager}. Consequently, the recent discovery of braid topology in non-Hermitian systems without symmetry has further sparked exploration in various fields including acoustics~\cite{ObservationAcousticNonHermitian2023zhang,ExperimentalCharacterizationThreeband2023zhang}, photonics~\cite{TopologicalComplexenergyBraiding2021wang}, and condensed matter physics~\cite{ExperimentalUnsupervisedLearning2022yu}.

\begin{figure}
	\centering
	\includegraphics[width=1.0\columnwidth]{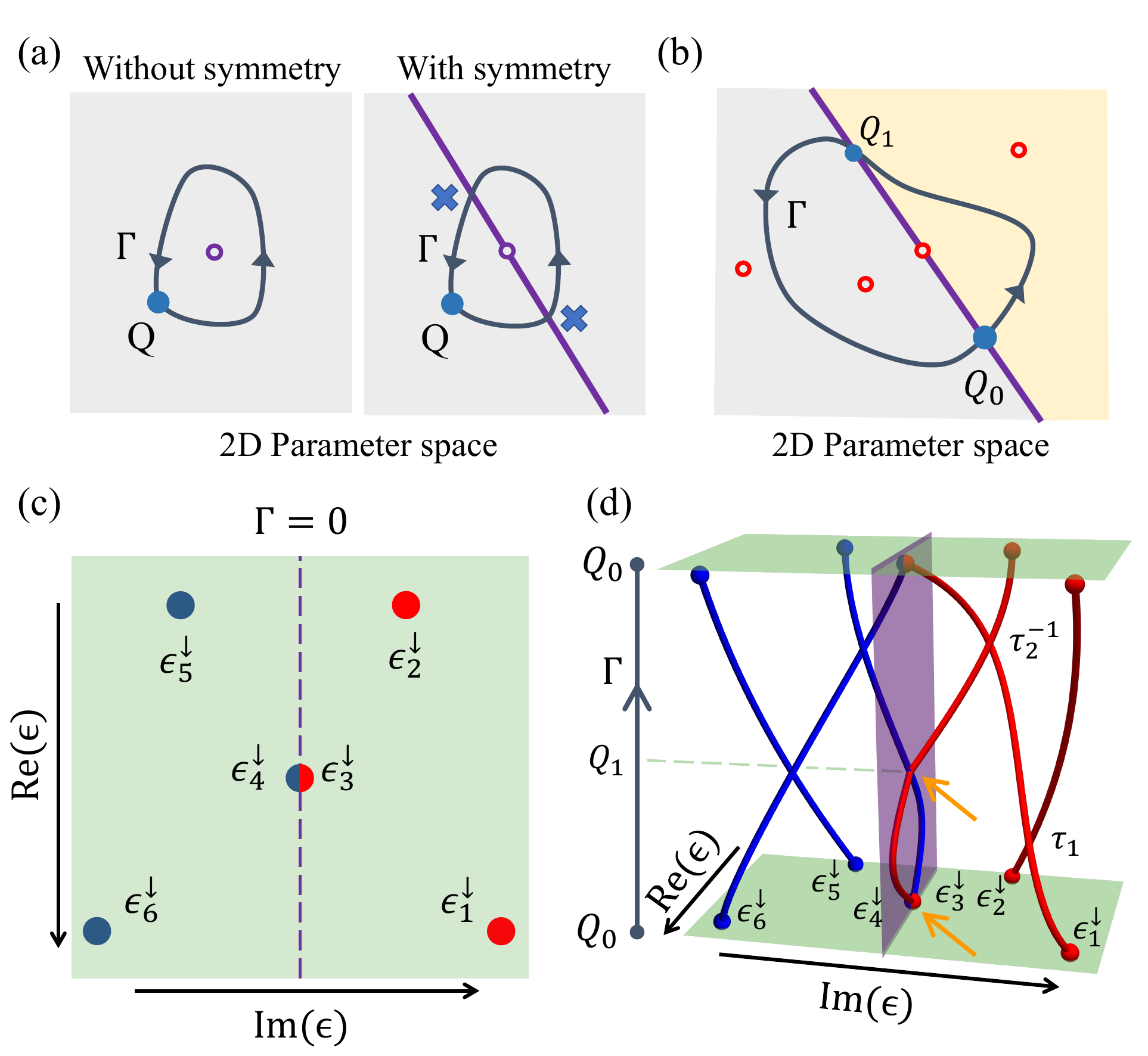}\\
	\caption {(a) A degeneracy point (the purple circle in the left panel) without symmetry becomes a nodal line (the purple line in the right panel) in the presence of certain symmetries. (b) A closed path  (the blue-black arrowed line) originating from $\mathrm{Q}_0$ enclosing two CPDPs (red circles) and passing through a nodal line in a 2D parameter space. This nodal line separates two areas (beige and gray) with probably different numbers of complex conjugate pairs of eigenvalues. (c) An illustration of sorting eigenvalues of a $6$ by $6$ non-Hermitian system with $\mathrm{PT}$ or $\mathrm{psH}$ symmetry at the beginning of the path $\Gamma$. (d) An illustration of braiding of the eigenvalue strands along the path $\Gamma$ in (b). $\tau_i$ denotes the braid algebra.}
 \label{fig1}
\end{figure}

Symmetry plays a vital role in topological phases. In Hermitian systems, the 10-fold Altland-Zirnbauer symmetry~\cite{ClassificationTopologicalInsulators2008schnyder} unlocks the classification of symmetry-protected topological phases beyond the scenarios without any symmetry. Similarly, in non-Hermitian systems, 38-fold symmetry enriches the classifications of wavefunction topology based on K-theory~\cite{SymmetryTopologyNonHermitian2019kawabata,NonHermitianTopologyExceptionalpoint2022ding}. Symmetries also have important consequences in the braid properties of spectral topology in non-Hermitian systems. Of particular importance are symmetries that can reduce the codimension of degeneracy points, such as pseudo-Hermiticity ($\mathrm{psH}$), parity-time symmetry ($\mathrm{PT}$), chiral symmetry ($\mathrm{CS}$), and parity-particle-hole symmetry ($\mathrm{CP}$)~\cite{DegeneraciesSymmetryBreaking2023melkani,SymmetryProtectedMultifoldExceptional2021delplace,ClassificationExceptionalNodal2021stalhammar,SymmetryprotectedExceptionalNodal2022sayyad,SymmetryprotectedTopologicalExceptional2023zhang,SymmetryHigherOrderExceptional2021mandal,EnhancedSensitivityHigherorder2017hodaei,NonlinearityenabledHigherorderExceptional2023bai,NonlinearExceptionalPoints2023bai,EnhancedSensingNondegraded2019xiao,ExceptionalpointbasedAccelerometersEnhanced2022kononchuk,CoherentPerfectAbsorption2021wang,SymmetryProtectedMultifoldExceptional2021delplace,DegeneraciesSymmetryBreaking2023melkani,SymmetryprotectedTopologicalExceptional2023zhang,ClassificationExceptionalNodal2021stalhammar,SymmetryprotectedExceptionalNodal2022sayyad,ProtectionAllNondefective2022sayyad,SymmetryHigherOrderExceptional2021mandal,NonHermitianSwallowtailCatastrophe2023hu}.  Such symmetries can significantly affect the braid topology. Specifically, for a system without symmetry, the generic degeneracy point has codimension 2 and thus occurs as a isolated point in a 2D parameter space [the left panel of Fig.~\ref{fig1}(a)]. The topology of such a point can be characterized by the closed path encircling it, based on the homotopy theory. However, with the above symmetries, the generic degeneracy has codimension 1~\cite{SymmetryProtectedMultifoldExceptional2021delplace,RealizingExceptionalPoints2022sayyad} and thus forms a nodal line in a 2D parameter space [the right panel of Fig.~\ref{fig1}(a)]. Consequently, its topology can no longer be characterized by a closed path as the path would unavoidably cross the singularities~\cite{NonHermitianSwallowtailCatastrophe2023hu,TopologicalClassificationIntersection2022jia}, disallowed in homotopy theory~\cite{AlgebraicTopology2002hatcher,NonHermitianSwallowtailCatastrophe2023hu,HomotopicalCharacterizationNonHermitian2021li,HomotopyCharacterizationNonHermitian2020wojcik,ExceptionalNonAbelianTopology2023guo}. As a result, the braid topology classifying degeneracy points in non-Hermitian systems with symmetry remains elusive.

In this work, we address this question by providing a general theory to elucidate the braid topology associated with symmetry-protected degeneracy points. We demonstrate that in the situation with reduced codimension, the eigenvalue topology can be characterized by the braid group $\mathcal{B}_m$, where $m$ is no longer the dimension of the system Hamiltonian. Furthermore, a specific type of degeneracy points, distinct from the ordinary degeneracy points in systems without symmetry, contributes to the braid topology. Additionally, a NACR governs the parametric evolution of these degeneracy points. We illustrate this NACR with exemplary systems. Our work not only extends the scope of the braid topology to non-Hermitian systems with symmetries, but also enables researchers to harness and manipulate these symmetry-protected degeneracy points in various physical domains, such as circuit systems~\cite{NonlinearityenabledHigherorderExceptional2023bai,ObservationAntiPTsymmetricExceptional2018choi,EnhancedSensingNondegraded2019xiao,ExceptionalpointbasedAccelerometersEnhanced2022kononchuk}, acoustic cavities~\cite{ObservationAcousticNonHermitian2023zhang,ExperimentalCharacterizationThreeband2023zhang}, coupled ring resonators~\cite{EnhancedSensitivityHigherorder2017hodaei,CoherentPerfectAbsorption2021wang,ParityTimeSymmetry2019ozdemir}.

{\color{blue}\textit{Theory.—}}
We introduce the notations used in this letter. $\mathbb{Z}$, $\mathbb{R}$, $\mathbb{R}_{+}$, and $\mathbb{C}$ denote the sets of integers, real numbers, non-negative real numbers, complex numbers, respectively. $\mathbb{C}_{+(-)}=\{x\in \mathbb{C} \mid \text{Im}(x)\ge (\le) 0\}$ while $\mathbb{C}_{+}^{0}=\{x\in \mathbb{C} \mid \text{Im}(x)> 0\}$. Let $(a_1,\ldots ,a_n)$ and $[a_1,\ldots ,a_n]$ be an ordered and unordered list of $n$ elements allowing repetition, respectively. $(a_1^{\downarrow},\ldots ,a_n^{\downarrow})$ stands for the ordered list obtained by sorting $[a_1,\ldots ,a_n]$, $a_i\in\mathbb{C}$ such that $\text{Im}(a_i^{\downarrow})>\text{Im}(a_j^{\downarrow})$ [and $\text{Re}(a_i^{\downarrow})\ge\text{Re}(a_j^{\downarrow})$ if $\text{Im}(a_i^{\downarrow})=\text{Im}(a_j^{\downarrow})$] for $i<j$. We denote $\mathrm{Conf}_{n}(\mathbb{F})$ ($\mathrm{UConf}_{n}(\mathbb{F})$) as the $n$-th ordered (unordered) configuration space of topological space $\mathbb{F}$ ($\mathrm{Conf}_{n}(\mathbb{F})=\{(m_1,\ldots ,m_n)\in\mathbb{F}^n\mid m_i\neq m_j \text{ for all } i \neq j\}$). $\mathcal{B}_n$ and $S_n$ denote the braid group on $n$ strands and the $n$ order symmetric group, respectively. $\sigma_i$ represents Pauli matrix.

Consider a $\mu \times \mu$ Hamiltonian $H(\lambda)$ that depends on $d$ dimensional parameters $\lambda\in\mathbb{R}^d$. We focus on the following antiunitary symmetries :
\begin{equation}\label{eq:pts}
\begin{aligned}
&\text { PT:} \quad  U_{\mathrm{PT}} H^{\ast}(\lambda) U_{\mathrm{PT}}^{-1}=H(\lambda) \quad  U_{\mathrm{PT}} U_{\mathrm{PT}}^{\ast}=\pm 1, \\
&\text { psH: } \quad  G_{\mathrm{psH}} H^{\dagger}(\lambda) G_{\mathrm{psH}}^{-1}=H(\lambda), \\ 
&\text { CP:} \quad  U_{\mathrm{CP}} H^{\ast}(\lambda) U_{\mathrm{CP}}^{-1}=-H(\lambda) \quad  U_{\mathrm{CP}} U_{\mathrm{CP}}^{\ast}=\pm 1, \\
& \text { CS: } \quad  G_{\mathrm{CS}} H^{\dagger}(\lambda) G_{\mathrm{CS}}^{-1}=-H(\lambda),  
\end{aligned}
\end{equation}
where $U_{s}$ ($G_{s}$) are unitary (Hermitian) matrices, and $\ast$ and $\dagger$ denote complex conjugate and conjugate transpose, respectively. Based on these symmetries, we study the $m$-fold degeneracy points, which correspond to the $m$-fold multiple roots of the characteristic polynomial of  $H(\mathbf{\lambda})$:
\begin{equation} 
  \label{eq:generalchaP}
  \begin{split}
      P_\lambda(E) &= \operatorname{det}[H(\lambda)-E] \\
  &=a_{\mu}(\lambda) E^{\mu}+\cdots +a_1(\lambda) E+a_0(\lambda).
  \end{split}
\end{equation}
The symmetries in Eq.~\eqref{eq:pts} require that all the coefficients of Eq.~\eqref{eq:generalchaP} $a_i$ are either real or imaginary. Thus, the codimension of a $2$-fold degeneracy point equals 1 in general~\cite{SymmetryProtectedMultifoldExceptional2021delplace} (see the Supplementary Materials Sec.~I \cite{SM} for details). Consequently, it is generally impossible for a closed path in a 2D parameter space to enclose a degeneracy point without encountering any other degeneracy points, as illustrated in the right panel of Fig.~\ref{fig1}(a) where the path (blue-black arrowed line) unavoidably passes through a line of degeneracy (purple line). Previous studies have introduced the winding number of the resultant vector to address this issue~ 
\cite{SymmetryProtectedMultifoldExceptional2021delplace}. However, the winding number is abelian; thus, it generally cannot capture the non-abelian topology intrinsic to non-Hermitian systems, although there may be exceptional case where the topology is abelian. 

In the main text, we primarily focus on $\mathrm{PT}$ symmetry. Other symmetries in Eq.~\eqref{eq:pts} are investigated in the Supplementary Materials Sec.~II \cite{SM}. The approach is summarized as follows: $\mathrm{CP}$ symmetry and $\mathrm{CS}$ symmetry can be mapped onto $\mathrm{PT}$ symmetry and $\mathrm{psH}$ symmetry, respectively, by transforming $H$ to $i H$. Furthermore,  $\mathrm{psH}$ symmetry can be encompassed in the subsequent discussion of $\mathrm{PT}$ symmetry.

For a Hamiltonian with $\mathrm{PT}$ symmetry, its eigenvalues are real or appear in complex conjugate pairs. We begin by assuming that within a parameter region, the number of conjugate pairs of eigenvalues remains constant and equals $m$. In this case, the topological space of eigenvalues can be represented as:
\begin{equation}
  \label{eq:pteigenlist}
  X^{(m)}=\{\left [ \epsilon_1,\dots, \epsilon_m ,\epsilon_m^{\ast} ,\dots, \epsilon_1^{\ast},\tilde{\epsilon} _1,\dots,\tilde{\epsilon} _{\mu-2m}\right]\},
  \end{equation}
  where $\text{Im}(\epsilon_i) \ge 0$, $\tilde{\epsilon}_i\in \mathbb{R}$, $\tilde{\epsilon}_i \neq \tilde{\epsilon}_j$ for all $i\neq j$. The unordered nature of the list arises from the equivalence of polynomials under the permutation of roots, while the condition $\tilde{\epsilon}_i \neq \tilde{\epsilon}_j$ results from the assumption that $m$ remains constant. It is important to note that the length of the eigenvalue list does not change at exceptional points according to our definition~\cite{NonHermitianPhysics2020ashida}. Next, we identify the singularity within the eigenvalue space. We define the complex conjugate pair degeneracy point (CPDP) as a degeneracy 
point where $\epsilon_{i}=\epsilon_{j}$ and simultaneously there exist another two eigenvalues $ \epsilon_{i}^{\ast}=\epsilon_{j}^{\ast}$ for $(i\neq j)$. The appearance of CPDP is a codimension 2 phenomenon (see details in the Supplementary Materials Sec.~III \cite{SM}).
To remove these singularities, we denote the space punctured by CPDPs as:
\begin{equation}
    \label{eq:eigenspacept1}
    X^{(m)}_0=\{\left(\epsilon_1^{\downarrow}, \ldots, \epsilon_m^{\downarrow}, (\epsilon_m^{\downarrow})^{\ast}, \ldots, 
(\epsilon_1^{\downarrow})^{\ast},\tilde{\epsilon} _1^{\downarrow},\dots,\tilde{\epsilon} _{\mu-2m}^{\downarrow}\right)\},
\end{equation}
where $\epsilon_i^{\downarrow}\neq\epsilon_j^{\downarrow}$ for $i\neq j$ and we have sorted the unordered eigenvalue list in Eq.~\eqref{eq:pteigenlist}, resulting in a unique representation. We define a map $g$ from $X^{(m)}_{0}$ to $\text{Conf}_m(\mathbb{C}_{+}) \times \text{Conf}_{\mu-2m}(\mathbb{R})$ as follows:
\begin{equation}
\begin{aligned}
        g&\left(\left(\epsilon_1^{\downarrow}, \ldots, \epsilon_m^{\downarrow}, (\epsilon_m^{\downarrow})^{\ast}, \ldots, 
(\epsilon_1^{\downarrow})^{\ast},\tilde{\epsilon} _1^{\downarrow},\dots,\tilde{\epsilon} _{\mu-2m}^{\downarrow}\right)\right) \\ &\equiv \left(\epsilon_1^{\downarrow}, \ldots, \epsilon_m^{\downarrow},\tilde{\epsilon} _1^{\downarrow},\dots,\tilde{\epsilon} _{\mu-2m}^{\downarrow}\right).
\end{aligned}
\end{equation}
The map $g$ is injective and continuous, and its inverse is also continuous. Therefore, it is a homeomorphism between $X^{(m)}_{0}$ and its image $g(X^{(m)}_{0})$:
\begin{equation}
  \begin{aligned}
    X^{(m)}_{0}&\cong g(X^{(m)}_{0}) \\ &= \{\left(\epsilon_1^{\downarrow}, \ldots, \epsilon_m^{\downarrow},\tilde{\epsilon} _1^{\downarrow},\dots,\tilde{\epsilon} _{\mu-2m}^{\downarrow}\right) \}\\
    &=\{\left(\left[\epsilon_1, \ldots, \epsilon_m\right],\left[\tilde{\epsilon} _1,\dots,\tilde{\epsilon} _{\mu-2m}\right]\right)\}\\
    &=\text{UConf}_{m}(\mathbb{C}_{+})\times  \text{UConf}_{\mu-2m}(\mathbb{R}).
\end{aligned}
\end{equation}
Consequently, we obtain the fundamental group of the punctured eigenvalue space $X_{0}^{(m)}$ (omitting the base point notation $Q$) as:
\begin{equation}
\label{eq:fhgpt1}
\begin{aligned}
    \pi_1(X_{0}^{(m)}) &=\pi_1\left(\text{UConf}_{m}(\mathbb{C}_{+})\times\text{UConf}_{\mu-2m}(\mathbb{R})\right)\\
    &=\pi_1\left(\text{UConf}_{m}(\mathbb{C}_{+})\right)\\
    &=\mathcal{B}_{m},
\end{aligned}
\end{equation}
where we use the fact that $\pi_1\left(\text{UConf}_{\mu-2m}(\mathbb{R})\right)$ is trivial~\cite{CollisionfreeMotionPlanning2017zapata,ConfigurationSpacesAlgebraic2018knudsen} and $\pi_1\left(\text{UConf}_{m}(\mathbb{C}_{+})\right)=\mathcal{B}_{m}$~\cite{RemarksHomotopyEquivalence2020plachta}. Therefore, the braid group $\mathcal{B}_{m}$ can be utilized to capture the eigenvalue topology, with its order being equal to the number of conjugate pairs of eigenvalues. 

For the $\mathrm{PT}^2=-1$ $(U_{\mathrm{PT}}U_{\mathrm{PT}}^{\ast}=-1)$ case, we have:
\begin{equation}
\langle \psi _r|U_{\mathrm{PT}}\mathcal{K} |\psi _r\rangle =-\langle \psi _r|U_{\mathrm{PT}}\mathcal{K} |\psi _r\rangle=0. 
\end{equation}
Here $|\psi _r\rangle$ denotes a right eigenvector with eigenvalue $\epsilon$, and $\mathcal{K}$ is the complex conjugate operator. This equation indicates that $|\psi _r\rangle$ and $U_{\mathrm{PT}}\mathcal{K} |\psi _r\rangle$ are linearly independent \cite{MatrixAnalysis2017horn}. Since $U_{\mathrm{PT}}\mathcal{K} |\psi _r\rangle$ is also an eigenvector with eigenvalue $\epsilon^{\ast}$, we conclude that $m=\mu/2$ $(m\in \mathbb{Z})$, and the number of complex conjugate pairs is equal to $\mu/2$. 

For the $\mathrm{PT}^2=1$ case (which also includes the $\mathrm{psH}$ case. See Supplementary Materials Sec.~II \cite{SM}), the number of complex conjugate pairs of eigenvalues (roots of the characteristic polynomial $P_{\lambda}(E)$) can vary. This number for a $\mu$-order real coefficient polynomial $P_{\lambda}(E)$ can be characterized by $\mathbb{Z}_2^{\mu}$, the revised sign list of its discriminant sequence~\cite{CompleteDiscriminationSystem1999liang,CompleteDiscriminationSystem1996yang}. 

The revised sign list can be introduced as follows: First, we consider a polynomial $p$'s  discrimination matrix $\operatorname{Discr}(p)$. This matrix is a variant of the Sylvester matrix, defined in Eq.~S18 in the Supplementary Materials Sec.~III \cite{SM}. We denote the determinant of the submatrix of $\operatorname{Discr}(p)$ formed by the first $2k$ rows and the first $2k$ columns as $D_{k}$ for $k=1,\cdots,n$. The resulting $n$-tuple
\begin{equation}
\label{eq:dlist}
    (D_1,D_2,\cdots,D_n)
\end{equation}
is referred as the discriminant sequence of the polynomial $p(x)$.
Then, the corresponding sequence 
\begin{equation}
    (\operatorname{sign}(D_1),\operatorname{sign}(D_2),\cdots,\operatorname{sign}(D_n))
\end{equation}
is termed the sign list of the discriminant sequence. Given a sign list $(s_1,s_2,\cdots,s_n)$, we construct a new list $(\Upsilon _1,\Upsilon _2,\cdots,\Upsilon _n)$, namely the \textbf{revised} sign list, as follows: 
\begin{itemize}
    \item If a section of the given list  $(s_i,s_{i+1},\cdots,s_{i+j})$ meets the condition where $s_i\neq 0$, $s_{i+1}= s_{i+2}=\cdots=s_{i+j-1}=0$, $s_{i+j}\neq 0$, then we replace $(s_{i},s_{i+1},\cdots,s_{i+j})$ with
$$
    (s_i,-s_i,-s_i,s_i,s_i,-s_i,-s_i,s_i,s_i,-s_i,\cdots).
$$
Specifically, let $\Upsilon _{i+r}=(-1)^{\text{floor}(\frac{r+1}{2})}\cdot s_i$ for $r=1,2,\cdots,j-1$. Otherwise, $\Upsilon _k=s_k$. For example, the revision of the sign list $(+,-,0,0,+)$ is $(+,-,+,+,+)$, where $0$s are replaced.
\end{itemize} 

Now, with revised sign list introduced, the number of complex conjugate pairs can be analyzed: it equals the number of sign changes in this revised sign list (see the Theorem~S1 in the Supplementary Materials Sec.~III \cite{SM} and Table~\ref{table:mt}). For a region with a constant number $m$ of conjugate pairs, we can apply directly the above analysis and use the braid group $\mathcal{B}_{m}$ to capture the eigenvalue topology. 

\begin{table}[h!]
  \caption{The number of complex conjugate pairs of eigenvalues (middle column) of a $6 \times 6$ non-Hermitian matrix exhibiting $\mathrm{PT}$ or $\mathrm{psH}$ symmetry, and the corresponding possible revised sign list of the discriminant sequence (right column) is shown. The number of complex conjugate pairs equals the number of sign changes in the sign list.}
  \label{table:mt}
  \centering
  \begin{tabular}{c|c|c}
  Degree $n$& Number of complex  & Possible revised sign list \\ & conjugate pairs & \\

  \hline
  \hline
  \multirow{4}*{$6$} & $3$  & $\left(+,-,+,-,-,-\right) $   \\
&  $2$  &   $\left(+,-,+,+,+,+\right) $\\
&$1$  &  $\left(+,-,-,-,-,-\right)$  \\
&$0$  &  $\left(+,+,+,+,+,+\right)$
  \end{tabular}
 \end{table}

We proceed to consider the situation where the number of complex conjugate pairs $m$ varies. For simplicity, we assume that $\mu=2n$ and consider the eigenvalue space where $m=n$ or $m=n-1$ denoted as
\begin{equation}
  X^{(n,n-1)}=\{[\epsilon_1,\ldots,\epsilon_{n-1},\epsilon_{n-1}^{\ast},\ldots,\epsilon_1^{\ast},\epsilon_n,\hat{\epsilon}_n]\},
\end{equation}
where when $m=n$, $\hat{\epsilon}_n=\epsilon_n^{\ast}$, $\text{Im}(\epsilon_i)\ge0$; when $m=n-1$, $\text{Im}(\hat{\epsilon}_n)=\text{Im}(\epsilon_n)=0$, $\text{Im}(\epsilon_i)> 0$ and $\text{Re}(\hat{\epsilon}_n)\neq\text{Re}(\epsilon_n)$. So, we have $\epsilon_n\in\mathbb{C}_{+}$ and $\hat{\epsilon}_n\in\mathbb{C}_{-}$. We define the following set in $X^{(n,n-1)}$:
\begin{equation}
\begin{alignedat}{3}
        \hat{X}^{(n-1)}&=\{&&[\epsilon_1,\ldots,\epsilon_{n-1},\epsilon_{n-1}^{\ast},\ldots,\epsilon_1^{\ast},\tilde{\epsilon}_n,\hat{\tilde{\epsilon}}_n]\\&\quad &&\mid \text{Im}(\epsilon_i)>0, \tilde{\epsilon}_n\in\mathbb{R},\hat{\tilde{\epsilon}}_n\in\mathbb{R}\}\\
        &=&&((\mathbb{C}_{+}^{0})^{n-1}/S_{n-1})\times (\mathbb{R}^2/S_{2}).
\end{alignedat}
\end{equation}

The eigenvalue space $X^{(n,n-1)}$ is the union of $X^{(n)}$ [defined in Eq.~\eqref{eq:pteigenlist}], and $\hat{X}^{(n-1)}$. And $X^{(n)}\cap \hat{X}^{(n-1)}=\{[\epsilon_1,\ldots,\epsilon_{n-1},\epsilon_{n-1}^{\ast},\ldots,\epsilon_1^{\ast},\tilde{\epsilon}_n,\tilde{\epsilon}_n] \mid \text{Im}(\epsilon_i)>0, \text{ }\tilde{\epsilon}_n\in\mathbb{R}\}=((\mathbb{C}_{+}^{0})^{n-1}/S_{n-1})\times \mathbb{R}$. We remove CPDPs from the eigenvalue space (denoted with the subscript $0$), resulting in $X_{0}^{(n)}=\text{UConf}_{n}(\mathbb{C}_{+})$, $\hat{X}^{(n-1)}_{0}=\text{UConf}_{n-1}(\mathbb{C}_{+}^{0})\times(\mathbb{R}^2/S_{2})$ and $X^{(n,n-1)}_{0}=X_{0}^{(n)}\cup \hat{X}_{0}^{(n-1)}$. These sets are open and path-connected due to the half-disk topology. By Seifert–Van Kampen theorem~\cite{AlgebraicTopology2002hatcher}, the fundamental group of $X^{(n,n-1)}_{0}$, with $Q_0\in X^{(n)}_{0}\cap \hat{X}^{(n-1)}_{0}$ as the base point, is isomorphic to the free product of the fundamental group of $X_{0}^{(n)}$, $\hat{X}^{(n-1)}_{0}$ with amalgamation of $\pi_{1}(X^{(n)}_{0}\cap \hat{X}^{(n-1)}_{0},Q_0)$:
\begin{equation}
\label{eq:vankampen}
\begin{aligned}
        &\pi_{1}(X^{(n,n-1)}_{0},Q_0)  \\&=\pi_{1}(X^{(n)}_{0},Q_0)*_{\pi_{1}(X^{(n)}_{0}\cap \hat{X}^{(n-1)}_{0},Q_0)}\pi_{1}(\hat{X}^{(n-1)}_{0},Q_0) \\ &=\mathcal{B}_{n},
\end{aligned}
\end{equation}
where $*_{\pi_{1}(X^{(n)}_{0}\cap \hat{X}^{(n-1)}_{0},Q_0)}$ denotes the amalgamation. The detailed proof of this result can be found in the Supplementary Materials Sec.~IV \cite{SM}. Therefore, we can analyze the eigenvalue topology for $\mu=2n$ and $m\in\{n,n-1\}$. The generalization to $\mu=2n+1$ is detailed in the Supplementary Materials Sec.~V \cite{SM}, where the order of braid group is also $n$. Thus, these two situations can be summarized as follows: when $m$ varies between $\text{floor}(\mu/2)$ and $\text{floor}(\mu/2)-1$, the order of braid group $n$ equals $\text{floor}(\mu/2)$.

In summary, CPDPs are classified by the braid group $\mathcal{B}_{n}$, where the order $n$ corresponds to the number of complex conjugate pairs of eigenvalues $m$ for both the cases of $\mathrm{PT}^2=\pm1$ and $\mathrm{psH}$. However, in the case of $\mathrm{PT}^2=1$ and the corresponding $\mathrm{psH}$ case with $m$ ranging from $\text{floor}(\mu/2)$ to $\text{floor}(\mu/2)-1$, the order $n$ is $\text{floor}(\mu/2)$. We note that our results are applicable to parameter spaces with dimensions higher than two where CPDPs are manifest as lines.

\begin{figure}
	\centering
	\includegraphics[width=1.0\columnwidth]{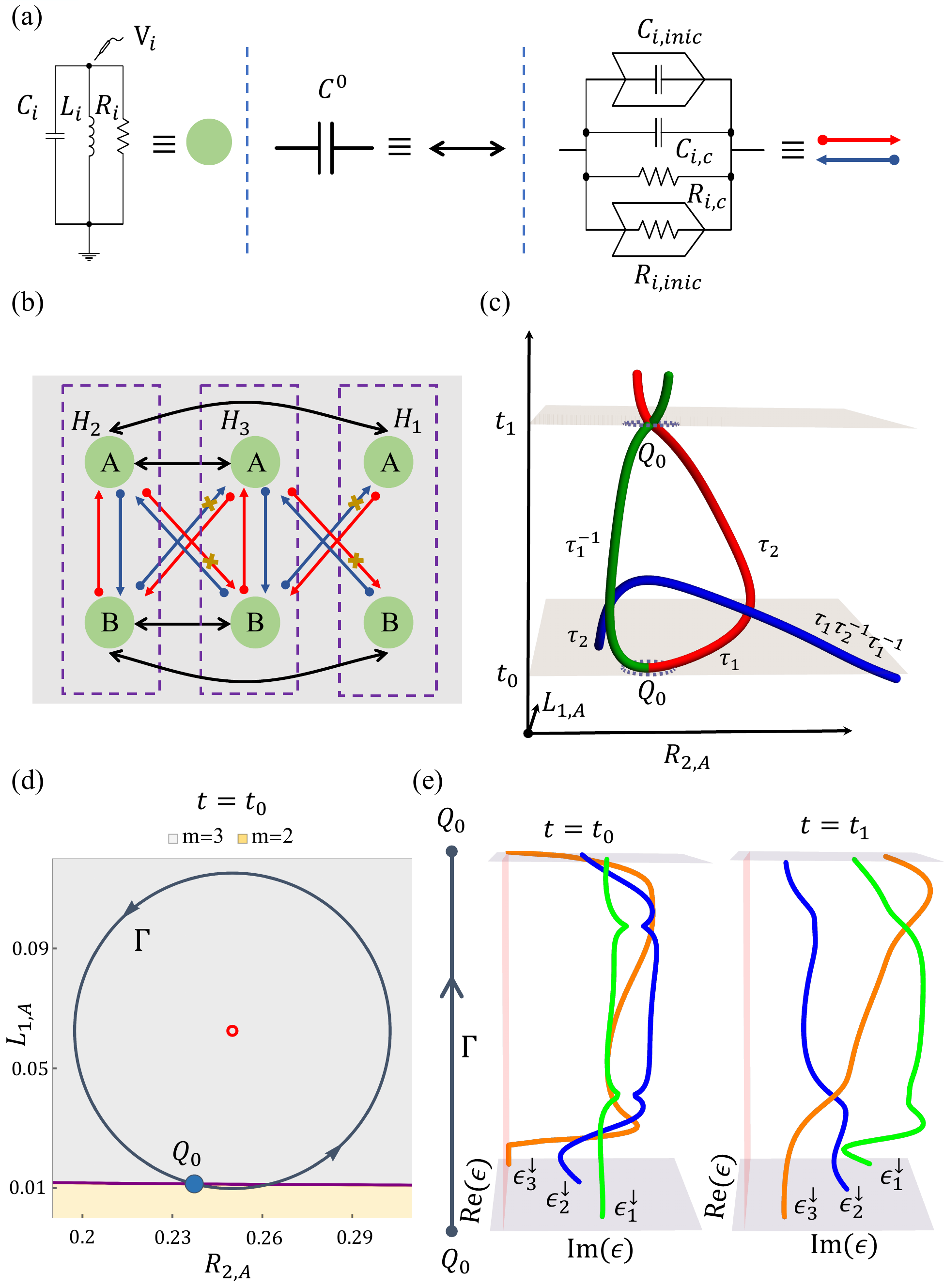}\\
	\caption {(a) Electric elements used in our model: a LC-R resonant cavity as a state, the capacitor $C^0$ as an identical coupling, and four independently tunable elements for arbitrary coupling. $C_{i,\text{inic}}$ and $R_{i,\text{inic}}$ represent a negative impedance converter with current inversion (INIC) associated with a capacitor and a resistor~\cite{GeneralizedBulkBoundary2020helbig,ChiralVoltagePropagation2019hofmann,NonHermitianSwallowtailCatastrophe2023hu}, respectively. (b) The circuit, where the subsystem enclosed by the purple dashed box is equivalent to $H_i$ in Eq.~\eqref{eq:pt1hi}. (c) Path-dependent annihilation of CPDP lines in the $(R_{2,A},L_{1,A},t)$ space. The blue-black dashed line denotes the path $\Gamma$ originating from point $Q_0$. (d) At $t=t_0$, the blue-black path $\Gamma$ enclosing the CPDP (the red open circle), traverses two regions (gray and beige) with different numbers of conjugate pairs of eigenvalues. (e) The first $3$ eigenvalue strands braiding along the path $\Gamma$ in (c) at $t=t_0$ and $t=t_1$. Here the green, blue, and orange lines represent eigenvalues with decreasing imaginary parts at the beginning of $\Gamma$. Detailed parameters are provided in Supplementary Materials Sec.~VII \cite{SM} and their projective trajectories are presented in Fig.~S3}.
	\label{fig2}
\end{figure}

The braid invariant characterizing CPDP or a path can be obtained using Artin braid word~\cite{BraidGroups2008kassel}. We consider a directional closed path with a base point, denoted as $\Gamma$ in Fig.~\ref{fig1}(b), which encloses two CPDPs (represented by red open circles) with a base point $\mathrm{Q}_0$ in a 2D parameter space. Initially, the eigenvalues are sorted as $(\epsilon_1^{\downarrow}, \epsilon_2^{\downarrow}, \dots, \epsilon_{\mu}^{\downarrow})$, as shown in Fig.~\ref{fig1}(c), and we focus on the first $n$ eigenvalues [represented by red dots in Fig.~\ref{fig1}(c)]. Importantly, as the eigenvalues evolve along $\Gamma$, we can consistently identify the eigenvalue strands originating from these $n$ eigenvalues as the first $n$ eigenvalues in the specified order (see Supplementary Materials Sec. II \cite{SM} for more details. The defective eigenspaces of exceptional points are utilized there to identify the eigenvalue strands).
For example, in Fig.~\ref{fig1}(d), we make such identifications as indicated by the orange arrows, resulting in the first $3$ eigenvalues being marked in red.
Subsequently, we sort these $n$ eigenvalues along $\Gamma$ and denote the crossing of the $i$-th eigenvalue over (under) the $i+1$-th eigenvalue as $\tau_i$ ($\tau_i^{-1}$), as shown in Fig.~\ref{fig1}(d). 
The braid invariant of the path $\Gamma$ is then given by the sequence of $\tau_i$ ($\tau_i^{-1}$) in the order they appear along $\Gamma$ [e.g. $\tau_1\tau_2^{-1}$ in  Fig.~\ref{fig1}(d)]. $\tau_i$ satisfies the braid relations~\cite{BraidGroups2008kassel} (see details in the Supplementary Materials Sec.~VI \cite{SM}).

{\color{blue}\textit{Examples.—}}
In this section, we present a model with $\mathrm{PT}$ symmetry and its circuit simulation to illustrate our theories. We confirm the occurrence of the path-dependent annihilation of CPDPs, which is brought about by the braid group, as well as an NACR that governs the behaviors of CPDPs. The NACR can be summarized as follows: for a time-varying Hamiltonian $H(\lambda(t))$ with braid topology, the braid invariants obtained from a fixed path with a fixed base point are conjugate between initial time $t_i$ and final time $t_f$, as long as there is no CPDP passing through the path during this time period. This can be expressed mathematically as:
\begin{equation}
\label{eq:nacr}
    b_{\Gamma}(t_f)=b_{\text{dyn}}^{-1}b_{\Gamma}(t_i)b_{\text{dyn}},
\end{equation}
where $b_{i}$ represents an element in the braid group and $b_{\text{dyn}}$ is a dynamical factor that acts indiscriminately~\cite{ExceptionalNonAbelianTopology2023guo,AlgebraicTopology2002hatcher,HomotopicalCharacterizationNonHermitian2021li,AlgebraicTopology2002hatcher}.

We consider a six-state model described by the following Hamiltonian:
\begin{equation}
  \label{eq:pt1h}
  H = \left(\begin{array}{ccc}
    H_1 & \mathbb{I}_{2} & 0 \\
    \mathbb{I}_{2} & H_2 & \Xi_1   \\
    \Xi_2 & \mathbb{I}_{2} & H_3
  \end{array}\right),
\end{equation}
where
\begin{equation}
  \label{eq:pt1hi}
  H_i = \left(\begin{array}{cc}
    \omega_i+i l_i & \kappa_i\\
    \kappa_i^{\ast} & \omega_i-i l_i
  \end{array}\right), 
\end{equation}
with $\omega_i,l_i \in \mathbb{R}$, $\kappa \in \mathbb{C}$. Additionally, we have $\Xi_1=\mathbb{I}_{2}+\xi(\sigma_1-\sigma_2)$ and $\Xi_2=
p_1\sigma_1+p_2\sigma_2$. The Hamiltonian $H$ in Eq.~\eqref{eq:pt1h} exhibits $\mathrm{PT}$ symmetry, where
\begin{equation}
  \hat{U}_{\mathrm{PT}}=\mathbb{I}_3\otimes \sigma_x,\quad \hat{U}_{\mathrm{PT}}\hat{U}_{\mathrm{PT}}^{\ast}=1.
\end{equation}

To realize this Hamiltonian, we employ the circuit system depicted in Fig.~\ref{fig2}. In this system, $\omega_{i,A(B)}$ represents the LC resonant frequency of the A(B) sublattice, and the complex voltage $V_{i,A(B)}$ at the node of the $A(B)$ resonators corresponds to the wave function at the $A(B)$ site [left panel of Fig.~\ref{fig2}(a)]. The gain or loss in each cavity, denoted by $l_i$, can be achieved using negative or positive resistors, while normal coupling is introduced with a capacitor [middle panel of Fig.~\ref{fig2}(a)]. Arbitrary nonreciprocal coupling is achievable with INICs [right panel of Fig.~\ref{fig2}(a)] \cite{NonHermitianSwallowtailCatastrophe2023hu,GeneralizedBulkBoundary2020helbig}. Therefore, the Hamiltonian in Eq.~\eqref{eq:pt1h} can be realized using the circuit shown in Fig.~\ref{fig2}(b) with appropriately chosen circuit elements (see detailed discussion in Supplementary Materials Sec.~VII \cite{SM}). For example, the coupling of the circuit inside the purple dashed box in Fig.~\ref{fig2}(b) is given by the matrix
\begin{equation}\setlength\arraycolsep{4pt}
\label{eq:circuit}
\begin{aligned}
     \begin{pmatrix}
\omega_{i,A}-\frac{i}{ 2 C_{i} R_{i,A} } & \wrapped { \frac{i\bigl(R_{i,\text{inic}}-R_{i,c}\bigr)}{2 C_{i}R_{i,\text{inic}}R_{i,c}} \\ {}+\omega_{i,B} \frac{C_{i,c}-C_{i,\text{inic}}}{2 C_{i}}} \\
\wrapped { \frac{i\bigl(R_{i,\text{inic}}+R_{i,c}\bigr)}{2 C_{i}R_{i,\text{inic}}R_{i,c}} \\ {}+\omega_{i,B} \frac{C_{i,c}+C_{i,\text{inic}}}{2 C_{i}}}  & \omega_{i,B}-\frac{i}{2 C_{i} R_{i,B}}
\end{pmatrix},
\end{aligned}
\end{equation}
which has the same form as $H_i$ in Eq.~\eqref{eq:pt1hi}.

This system exhibits the path-dependent annihilation of degeneracy points, a unique characteristic of systems with non-abelian topology~ \cite{ExceptionalNonAbelianTopology2023guo,NonAbelianBandTopology2019wu}. To observe this phenomenon, we allow certain circuit parameters to vary with time while preserving $\mathrm{PT}$ symmetry. Figure~\ref{fig2}(c) illustrates the evolution of CPDPs in the 3D $(R_{2,A},L_{1,A},t)$ parameter space. At $t_0$, a pair of CPDPs emerges [green and red lines in Fig.~\ref{fig2}(c)]. We associate the red CPDP with the braid word $b_{\text{red}}=\tau_1$ and the green CPDP with the braid word $b_{\text{green}}=\tau_1^{-1}$, both with respect to a base point $\mathrm{Q_0}$. On either side of these CPDPs, there are two additional blue CPDPs associated with the braid words $b_{\text{blue},l}=\tau_2$ and $b_{\text{blue},r}=\tau_1\tau_2^{-1}\tau_1^{-1}$ at $t_0$. As depicted in Fig.~\ref{fig2}(d), the path $\Gamma$ traverses two regions with $m=2$ and $m=3$ conjugate pairs at $t=t_0$. Accordingly, we select the base point $Q_0$ on the exceptional line [purple line in Fig.~\ref{fig2}(d)] that separates these two regions. As shown in Fig.~\ref{fig2}(c), the red and green CPDPs subsequently deviate from their original paths and encircle the nodal line formed by the blue CPDP along the $t$ axis. When the blue nodal line with the braid word $b_{\text{blue},r}$ passes above the red nodal line, the braid invariant of the red line becomes conjugate to the blue line: $\bar{b}_{\text{red}}=b_{\text{blue},r}^{-1}b_{\text{red}} b_{\text{blue},r}=\tau_2$. Additionally, there is a dynamical factor $b_{\text{dyn}}=\tau_1\tau_2\tau_1\tau_2$ involved in this time evolution. According to the NACR in Eq.~\eqref{eq:nacr}, the braid word of the path $\Gamma$ at $t=t_1$ is given by $b_{\Gamma} (t_1)=b_{\text{dyn}}^{-1}\bar{b}_{\text{red}}b_{\text{green}}b_{\text{dyn}}=\tau_2^{-1}\tau_1$. Consequently, instead of annihilation, a third-order CPDP appears when the red and green nodal lines merge at $t_1$ before subsequently splitting again. As illustrated in Fig.~\ref{fig2}(e), the braid words of the same path $\Gamma$ differ at $t=t_0$ and $t=t_1$: $1$ (left panel) and $\tau_2^{-1}\tau_1$ (right panel), respectively. (Detailed algebra can be found in Supplementary Materials Sec.~VI \cite{SM}.)

The example above verifies that CPDPs follow a NACR, even when the number of conjugate pairs undergoes changes. Consequently, the braid topology governing CPDPs grants us the ability to manipulate and harness these singularities. In comparison to the case without symmetry, the merge point at $t=t_1$ exhibits two third-order degeneracies, suggesting potential applications in sensing devices~\cite{EnhancedSensingNondegraded2019xiao,EnhancedSensitivityHigherorder2017hodaei,NonlinearExceptionalPoints2023bai,NonlinearityenabledHigherorderExceptional2023bai}. The model with $\mathrm{PT}^2=-1$ is provided in Supplementary Materials Sec. VIII \cite{SM}.

{\color{blue}\textit{Conclusion and Discussions.—}}
In conclusion, we have provided a systematic investigation of the braid topology in non-Hermitian systems where symmetries play a crucial role in reducing the codimension of degeneracy points.
Instead of relying solely on the oversimplified winding number topology~\cite{SymmetryProtectedMultifoldExceptional2021delplace}, we have uncovered the fascinating phenomenon of path-dependent coalescence of CPDPs.
The existence of the NACR under symmetries aligns with the non-abelian nature of non-Hermitian multiband eigenvalue topology \cite{HomotopyCharacterizationNonHermitian2020wojcik,HomotopicalCharacterizationNonHermitian2021li,ExceptionalNonAbelianTopology2023guo,KnotTopologyExceptional2022hu}, providing a more coherent picture. Moreover, the models we have presented can be experimentally realized in various platforms, such as acoustic cavities \cite{ObservationAcousticNonHermitian2023zhang,ExperimentalCharacterizationThreeband2023zhang}, optical waveguides \cite{DynamicallyEncirclingExceptional2016doppler,SymmetryRecoveryCoupled2015rivolta} and ring resonators \cite{TopologicalComplexenergyBraiding2021wang,GeneratingArbitraryTopological2021wang}. These findings empower researchers in diverse fields to harness symmetry in non-Hermitian systems, leading to significant implications and inspiring further investigations into symmetry-protected non-Hermitian degeneracy points and their applications.

\emph{Note added.---} We become aware of a parallel work \cite{HomotopySymmetryNonHermitian2023yang} which overlaps with parts of this work.

\begin{acknowledgments}
This work is supported by the National Key Research and Development Program of China (Grant No. 2022YFA1404900), the National Natural Science Foundation of China (Grant No. 12274330, 12274332), and the Knowledge Innovation Program of Wuhan-Shuguang (Grant No. 2022010801020125).
\end{acknowledgments}

\bibliography{ref}

\newpage
\onecolumngrid

\titlespacing{\subsection}{0pt}{7.5ex plus 1ex minus .2ex}{2em}
\renewcommand{\theequation}{S\arabic{equation}}
\renewcommand{\thefigure}{S\arabic{figure}}
\renewcommand{\thetable}{S\arabic{table}}
\renewcommand{\thethm}{S\arabic{thm}}
\setcounter{equation}{0}
\setcounter{figure}{0}
\setcounter{table}{0}

\begin{center}
    {\bf \large Supplementary Material for\\ ``Braiding topology of symmetry-protected degeneracy points in non-Hermitian system''}
\end{center}

\subsection{{I} The codimension of \texorpdfstring{$2$}{2}-fold degeneracy points}
In this section, we prove that the appearance of a $m$-fold degeneracy point in a parametric space $\bar{\lambda}$ is equivalent to requiring the resultant of $P_{\bar{\lambda}}(E)$ and $\sum^{m-1}_{i=1} q_{i} \frac{ \partial^{i} P_{\bar{\lambda}}(E)} {\partial E^{i}}$ equals zero regardless of the value of $q_{i}$. Thus, we can obtain the codimension of $2$-fold degeneracy points.

The resultant of two polynomials is defined as the determinant of the Sylvester matrix formed by their coefficients. Let $A(x)$ and $B(x)$ be two non-zero polynomials of degrees $d$ and $e$, respectively:
\begin{eqnarray}
\begin{aligned}
  &A(x)=a_d x^d + a_{d-1} x^{d-1}+\dots+a_{1} x^{1}+a_{0},\\
&B(x)=b_e x^e + b_{e-1} x^{e-1}+\dots+b_{1} x^{1}+b_{0}.
\end{aligned}
\end{eqnarray} 
Their resultant, denoted as $\operatorname {res} (A,B)$, is defined as:
\begin{eqnarray}
    \operatorname {res} (A,B)=a_{d}^{e}b_{e}^{d}\prod _{\begin{array}{c}1\leq i\leq d\\1\leq j\leq e\end{array}}(\alpha _{i}-\beta _{j}),
\end{eqnarray}
where $\alpha_i$, $\beta_i$ are respectively the roots of $A$ and $B$. It's clear that when two polynomials $A$ and $B$ have at least a common root the resultant $\operatorname {res} (A,B)=0$.
Let $P(x)$ be a polynomial of degree $n$ with root $\alpha_i$
\begin{eqnarray}
    P(x)=\prod_{i=1}^{n}(x-\alpha_i).
\end{eqnarray}
a $m$-fold degeneracy points (multiple-root) $a=\alpha_1=\alpha_2=\dots=\alpha_m$ requires that the polynomial can be written as follows:
\begin{eqnarray}
    P(x)=(x-a)^m\prod_{i=m+1}^{n}(x-\alpha_i).
\end{eqnarray}
So, the factorization of its $j$-th order derivative $\frac{ \partial^{j} P(x)} {\partial x^j}$ ($j=0\dots m-1$) is $(x-a)\bar{P}^{(j)}$, where $\bar{P}^{(j)}$ is a polynomial of $x$ . It follows that the resultants between $P(x)$ and its derivatives of order $0$ to $m-1$ all equal zero~\cite{DiscriminantsResultantsMultidimensional2008gelfand}.
This proves the sufficiency condition:
\begin{eqnarray}
    \sum^{m-1}_{j=1} q_{j} \frac{ \partial^{j} P(x)} {\partial x^{j}}=(x-a)\sum^{m-1}_{j=1} q_{j} \bar{P}^{(j)} \Longrightarrow  \operatorname {res} (P(x),\sum^{m-1}_{i=1} q_{i} \frac{ \partial^{i} P(x)} {\partial x^{i}})=0
\end{eqnarray}
For the necessity, the resultant of $\sum^{m-1}_{i=1} q_{i} \frac{ \partial^{i} P(x)} {\partial x^{i}}$ and  $ P(x)$ equaling zero implies that $P(x)$ and the combination of its derivatives $\sum^{m-1}_{i=1} q_{i} \frac{ \partial^{i} P(x)} {\partial x^{i}}$ have a common root $a(q)$. Here $a(q)$ could depend on $q$. However, since their resultant equals to zero regardless $q$, the common root $a(q)$ would be $a$, irrelevant to $q$:
\begin{eqnarray}
    \sum^{m-1}_{j=1} q_{j} \frac{ \partial^{j} P(x)} {\partial x^{j}}=(x-a)\sum^{m-1}_{j=1} q_{j} \bar{P}^{(j)}.
\end{eqnarray}
So, $j$-th order derivative of $ P(x)$ factorize by $x-a$. The polynomial $P(x)$ has this form:
\begin{eqnarray}
    P(x)=(x-a)^m\prod_{i=m+1}^{n}(x-\alpha_i),
\end{eqnarray}
from which we can see the $P(x)$ have a $m$-fold multiple root $a$.

Thus, for a real or imaginary coefficient polynomial $P(x)$, the appearance of $2$-fold root is equivalent to $\operatorname {res} (P(x),\frac{\partial P(x)}{\partial x})=0 $, which is one real equation (or can be transformed into one real equation). So, the codimension of $2$-fold root of a real or imaginary coefficient polynomial equals to $1$.

\subsection{{II} Generalization to other symmetries and details about eigenvalue strands}

In this section, we discuss the situations with other symmetries included in Eq.~(1) of main text. We also discuss how to define the first $n$ eigenvalue strands and outline the equivalence relationship.

\subsubsection*{Symmetries}
\paragraph{$\mathrm{CP}$ symmetry.}

Consider a Hamiltonian $H_{\mathrm{CP}}$ that satisfies $\mathrm{CP}$ symmetry:
    \begin{equation}
        U_{\mathrm{CP}} H_{\mathrm{CP}}^{\ast}(\lambda) U_{\mathrm{CP}}^{-1}=-H_{\mathrm{CP}}(\lambda),
    \end{equation}
we define a new Hamiltonian $H^{\prime}=i H_{\mathrm{CP}}$. Then it satisfies $\mathrm{PT}$ symmetry:
    \begin{equation}
        U_{\mathrm{CP}} (H^{\prime})^{\ast}(\lambda) U_{\mathrm{CP}}^{-1}=H^{\prime}(\lambda).
    \end{equation}
Thus, we can apply the theory developed in the main text to classify the degeneracy points in systems with $\mathrm{CP}$ symmetry.
\paragraph{$\mathrm{psH}$ symmetry.} Consider a Hamiltonian $H_{\mathrm{psH}}$ that satisfies $\mathrm{psH}$ symmetry:
\begin{equation}
    G_{\mathrm{psH}} H_{\mathrm{psH}}^{\dagger}(\lambda) G_{\mathrm{psH}}^{-1}=H_{\mathrm{psH}}(\lambda).
\end{equation}
Its characteristic polynomial satisfies:
\begin{equation}
\begin{aligned}
    P_\lambda(E) &= \operatorname{det}[H_{\mathrm{psH}}(\lambda)-E]
    =a_{\mu}(\lambda) E^{\mu}+\cdots a_1(\lambda) E+a_0(\lambda)\\
    &=\operatorname{det}[G_{\mathrm{psH}} H_{\mathrm{psH}}^{\dagger}(\lambda) G_{\mathrm{psH}}^{-1}-E]
    =\operatorname{det}[ H_{\mathrm{psH}}^{\ast}(\lambda)-E] 
    =a^{\ast}_{\mu}(\lambda) E^{\mu}+\cdots a^{\ast}_1(\lambda) E+a^{\ast}_0(\lambda).
\end{aligned}
\end{equation}
As a result, all $a_{i}$ are all real numbers. The classification of $\mathrm{psH}$ symmetry-protected degeneracy points can be encompassed in the $\mathrm{PT}^2=1$ case discussed in the main text, where we only use the condition that all coefficients are real \cite{PTsymmetryEntailsPseudoHermiticity2020zhang}.
\paragraph{$\mathrm{CS}$ symmetry.}
Consider a Hamiltonian $H_{\mathrm{CS}}$ that satisfies chiral symmetry:
\begin{equation}
    \quad G_{\mathrm{CS}} H_{\mathrm{CS}}^{\dagger}(\lambda) G_{\mathrm{CS}}^{-1}=-H_{\mathrm{CS}}(\lambda).
\end{equation}
we define a new Hamiltonian $H^{\prime}=i H_{\mathrm{CS}}$. Then it satisfies $\mathrm{psH}$ symmetry:
    \begin{equation}
         G_{\mathrm{psH}} (H^{\prime})^{\dagger}(\lambda) G_{\mathrm{psH}}^{-1}=(H^{\prime})(\lambda).
    \end{equation}
So, we can map a Hamiltonian with $\mathrm{CS}$ symmetry to a Hamiltonian with $\mathrm{psH}$ symmetry, and then the classification is straight forward. 

\subsubsection*{Identification of the first $n$ eigenvalue strands}
Let us assume that we sort eigenvalues in order at the beginning of a path $\Gamma$: $(\epsilon_1^{\downarrow}, \epsilon_2^{\downarrow}, \dots,
\epsilon_{\mu}^{\downarrow})$. Since the order of the braid group $n$ is greater than or equal to the number of conjugate pairs of eigenvalues, we have $\text{Im}(\epsilon_i^{\downarrow})\ge 0$ for $i\in\{1,2,\dots,n\}$ and $\text{Im}(\epsilon_i^{\downarrow})\le 0$ for $i\in\{n+1,n+2,\dots,\mu\}$. So, the swap between an eigenvalue $i\in\{1,2,\dots,n\}$ and an eigenvalue $j\in\{n+1,n+2,\dots,\mu\}$ can be divided into two situations:
\begin{itemize}
    \item The swap between eigenvalue $\epsilon_i$ with $\text{Im}(\epsilon_i) > 0$ and its complex conjugate $\epsilon_i=\epsilon_i^{\ast}$;
    \item The swap between eigenvalue $\epsilon_i$ with $\text{Im}(\epsilon_i) = 0$ and eigenvalue $\epsilon_j$ with $\text{Im}(\epsilon_j) = 0$ where $\text{Re}(\epsilon_i)>\text{Re}(\epsilon_j)$.
\end{itemize}
For the first situation, the intermediate value theorem guarantees that during the swap process, a point where $\text{Im}(\epsilon_i) = 0$ must be encountered. And at this point we have $\epsilon_j=\epsilon_i^{\ast}=\epsilon_i$, making it impossible to distinguish them. Thus, we can always identify that the eigenvalue strand of $\epsilon_i$ is among the first $n$ eigenvalues as we need to define the braiding topology. For the second situation, a point where $\epsilon_i=\epsilon_j$ is also guaranteed by the intermediate value theorem. Then, we can also always identify that the eigenvalue strand of $\epsilon_i$ is among the first $n$ eigenvalues.

\subsubsection*{A intuitive picture about the equivalence relationship}
Here we provide an intuitive explanation why the group action associated with equivalence relationship is described by the $S_n$ rather than $S_{\mu}$, where $n$ and $\mu$ denote the order of the braid group and the Hamiltonian. Take the situation where $\mathrm{PT}^2=-1$ for example. In the braiding process, at the beginning we order eigenvalues list $(\epsilon_1^{\downarrow}, \ldots  ,\epsilon_n^{\downarrow},(\epsilon_n^{\downarrow})^{\ast}, \ldots, (\epsilon_1^{\downarrow})^{\ast})$. We denote the generators of $S_{\mu}$, describing the possible permutation actions involved in this process, as follows:
\begin{equation} \label{eq:generators} \begin{cases} \varsigma _i =(\epsilon_i^{\downarrow},\epsilon_{i + 1}
^{\downarrow} )
\quad \text{swap } \epsilon_i^{\downarrow} \text{ and } \epsilon_{i+1}^{\downarrow},\text{
} i \le n - 1 \\ \varsigma 
= (\epsilon_n^{\downarrow},(\epsilon_n^{\downarrow})^{\ast} )\quad \text{swap } \epsilon_n^{\downarrow} \text{ and }
(\epsilon_n^{\downarrow})^{\ast} \\ \bar{\varsigma }_i = ((\epsilon_i^{\downarrow})^{\ast},
(\epsilon_{i + 1}^{\downarrow})^{\ast} )\quad  \text{swap }
(\epsilon_i^{\downarrow})^{\ast} \text{ and } (\epsilon_{i + 1}^{\downarrow})^{\ast}, i \le n - 1.  \end{cases} \end{equation} 

Here we make two observations to qualify permutation actions so that we are able to obtain well-defined group actions on $X_{0}^{n}$. Firstly, during the continuous process of swapping points, once $
\epsilon_i^{\downarrow}$ and $\epsilon_{i+1}^{\downarrow}$ are swapped, $
(\epsilon_i^{\downarrow})^{\ast}$ and $(\epsilon_{i+1}^{\downarrow})^{\ast}$ are also swapped. Consequently, in this situation, $\varsigma _i$ is accompanied by $\bar{\varsigma }_i$, and they commute 
$\varsigma _i \bar{\varsigma }_i=\bar{\varsigma }_i \varsigma _i$. Secondly, $\varsigma $ implies that at a certain point during the swapping process $
\text{Im}(\epsilon_n^{\downarrow})=0$, resulting in $\epsilon_n^{\downarrow}= 
(\epsilon_n^{\downarrow})^{\ast}$. This conclusion arises from the intermediate value theorem, which ensures the existence of such a point during the continuous process of swapping between  $\text{Im}(\epsilon)\ge 0$ and 
$\text{Im}(\epsilon)\le 0$. Given that at this point $\epsilon_n^{\downarrow}=
(\epsilon_n^{\downarrow})^{\ast}$, one cannot identify these two points, making it uncertain whether they exchange or not. Thus, here we adopt the convection that during this process, the two points do not swap $
\varsigma  \rightarrow 1$. Now, with these two observations in mind, the generators of the group associated with the equivalence relationship $G_e$ are $\varsigma _i\bar{\varsigma _i}\text{
} (i \le n - 1)$. 

The collection of generators $\varsigma _1\bar{\varsigma }_1,\varsigma _2\bar{\varsigma }_2,\cdots,\varsigma _{n-1}\bar{\varsigma }_{n-1}$ subject to the following relationship:

\begin{itemize}
    \item $ \varsigma _i\bar{\varsigma }_i \varsigma _i\bar{\varsigma }_i =(\varsigma _i)^2(\bar{\varsigma }_i)^2=1 $,
    \item $ \varsigma _i\bar{\varsigma }_i\varsigma _j\bar{\varsigma }_j =\varsigma _i\varsigma _j \bar{\varsigma }_i\bar{\varsigma }_j= \varsigma _j \bar{\varsigma }_j \varsigma _i \bar{\varsigma }_i, \quad \text{for }\mid i-j \mid >1$,
    \item $(\varsigma _i\bar{\varsigma }_{i}\varsigma _{i+1}\bar{\varsigma }_{i+1})^3=(\varsigma _i\varsigma _{i+1})^3(\bar{\varsigma }_{i}\bar{\varsigma }_{i+1})^3=1$,
\end{itemize}
where we use the commutation relation between $\varsigma _i$ and  $\bar{\varsigma }_j$.
We can see those group generators follows the same relationship as the generators of $S_n$ and have the same number of generators. Thus, we can define a isomorphic map $f(\varsigma _i\bar{\varsigma }_i)=\varsigma _i$ \cite{LieGroupsLie2015halla,GroupTheoryApplication2010dresselhausa}. The group generated by those generators is isomorphic to the symmetric group $S_n$ 
\begin{equation}
    G_e\cong S_n.
\end{equation}

\subsection{{III} Revised sign list of the discriminant sequence and the codimension of CPDPs }

\subsubsection*{Revised sign list of the discriminant sequence}
First, let $p$ be a non-zero polynomial of degree $n$,
\begin{eqnarray}
 p(x)=a_{n}x^{n}+\cdots+a_{2}x^{2}+a_{1}x^{1}+a_{0} .
\end{eqnarray}
Its derivative is:
\begin{eqnarray}
 p^{\prime}(x)=n a_{n}x^{n-1}+\cdots+2 a_{2} x^{1}+a_{1} .
\end{eqnarray}
A variant of the Sylvester matrix of $p(x)$ and $p^{\prime}(x)$, which is named the discrimination matrix, is defined as follows:
\begin{equation}
   \operatorname{Discr}(p)= \begin{pmatrix}
a_n & a_{n-1} & \cdots & a_1 & a_0 & 0 & 0 &\cdots & 0 & 0 \\
0 & na_{n} & \cdots & 2a_2 & a_1 & 0 & 0 & \cdots & 0 & 0 \\
0 & a_n & a_{n-1} & \cdots & a_1 & a_0 & 0  & \cdots & 0 & 0 \\
0 & 0 & na_{n} & \cdots & 2a_2 & a_1 &  0 &\cdots & 0 & 0  \\
0 & 0 & \cdots & \cdots & \cdots & 0 &0 & 0 & 0 & 0 \\
0 & 0 & \cdots & \cdots & \cdots & 0 &0 & 0 & 0 & 0\\
0 & 0  &\cdots &0 & 0& a_n & a_{n-1} & \cdots & a_1 & a_0\\
0 & 0 &\cdots & 0& 0& 0 &  na_{n} & \cdots & 2a_2 & a_1 \\
\end{pmatrix}.
\end{equation}
We use $D_{k}$ to denote the determinant of the submatrix of $\operatorname{Discr}(p)$ formed by the first $2k$ rows and the first $2k$ columns for $k=1,\cdots,n$. We call the resulting n-tuple
\begin{equation}
\label{eq:mdlist}
    (D_1,D_2,\cdots,D_n)
\end{equation}
the discriminant sequence of the polynomial $p(x)$. Then the corresponding sequence 
\begin{equation}
    (\operatorname{sign}(D_1),\operatorname{sign}(D_2),\cdots,\operatorname{sign}(D_n))
\end{equation}
is dubbed as the sign list of the discriminant sequence. Given a sign list $(s_1,s_2,\cdots,s_n)$, we construct a new list $(\Upsilon _1,\Upsilon _2,\cdots,\Upsilon _n)$, i.e., the \textbf{revised} sign list, as follows: 
\begin{itemize}
    \item If $(s_i,s_{i+1},\cdots,s_{i+j})$ is a section of the given list and $s_i\neq 0$, $s_{i+1}= s_{i+2}=\cdots=s_{i+j-1}=0$, $s_{i+j}\neq 0$, then we replace $(s_{i},s_{i+1},\cdots,s_{i+j})$ with
\begin{equation}
    (s_i,-s_i,-s_i,s_i,s_i,-s_i,-s_i,s_i,s_i,-s_i,\cdots),
\end{equation}
i.e. let $\Upsilon _{i+r}=(-1)^{\text{floor}(\frac{r+1}{2})}\cdot s_i$ for $r=1,2,\cdots,j-1$. Otherwise, let $\Upsilon _k=s_k$. For example, the revision of the sign list $(+,-,0,0,+)$ is $(+,-,+,+,+)$, where $0$s are replaced.
\end{itemize}

The following theorem reveals the relationship between the number of complex conjugate pairs and the revised sign list. 
\begin{thm}
\label{rootnumber}
    Given a polynomial $p(x)$ with real coefficients,
\begin{equation}
     p(x)=a_{n}x^{n}+\cdots+a_{2}x^{2}+a_{1}x^{1}+a_{0}:
\end{equation}
If the number of the sign changes of the revised sign list of  $(D_1(p),D_2(p),\cdots,D_n(p))$ is $m$, then the number of the pairs of distinct conjugate imaginary roots of $p(x)$ equals $m$; Furthermore, if the number of non-zero members of the revised sign list is $l$, then the number of the distinct real roots of $p(x)$ equals $l-m$.
\end{thm}
The detailed proof of Theorem~\ref{rootnumber} can be found in Refs.~\cite{CompleteDiscriminationSystem1996yang,CompleteDiscriminationSystem1999liang}. The second part of Theorem~\ref{rootnumber} includes the case when $a_n=0$ and then $l \neq n$. We take a $6 \times 6$ non-Hermitian Hamiltonian exhibiting $\mathrm{PT}$ or $\mathrm{psH}$ symmetry as an example for illustration. Because its characteristic polynomial is a real coefficient polynomial of degree 6, the number of conjugate pairs of eigenvalue can take values $\{0,1,2,3\}$ as shown in the middle column of Table~\ref{table:t}. If there are $3$ pairs of complex conjugate eigenvalues, then the sign changes three times for the revised sign list from left to right. A possible list can be $\left(+,-,+,-,-,-\right) $ as provided in the right column of Table~\ref{table:t}.

\begin{table}[h!]
  \caption{The number of complex conjugate pairs of eigenvalues (middle column) of a $6 \times 6$ non-Hermitian matrix exhibiting $\mathrm{PT}$ or $\mathrm{psH}$ symmetry [defined in Eq.~(1) in the main text] and the corresponding possible revised sign list of the discriminant sequence (right column). The number of complex conjugate pairs equals to the number of sign changes in the sign list.}
  \label{table:t}
  \centering
  \begin{tabular}{c|c|c}
  Degree $n$& Number of complex  & Possible revised sign list \\ & conjugate pairs & \\

  \hline
  \hline
  \multirow{4}*{$6$} & $3$  & $\left(+,-,+,-,-,-\right) $   \\
&  $2$  &   $\left(+,-,+,+,+,+\right) $\\
&$1$  &  $\left(+,-,-,-,-,-\right)$  \\
&$0$  &  $\left(+,+,+,+,+,+\right)$
  \end{tabular}
 \end{table}
 
\subsubsection*{The codimension of CPDP}
At each CPDP, we have two sets of two-fold multiple roots: $x_{i}=x_{j}$ and $x_{i}^{\ast}=x_{j}^{\ast}$. Thus, according to Sec. I, the appearance of a CPDP implies that there are two common roots between the characteristic polynomial $P(x)$ and its derivative $P^{\prime}(x)$. Thus, the degree of the polynomial greatest common divisor $\mathrm{deg}\left(\mathrm{gcd}(P(x),P^{\prime}(x))\right)\ge2$. The degree of greatest common divisor of two polynomials $P$ and $Q$ has a deep connection with the subresultants coefficients of these two polynomials $P$ and $Q$~\cite{AlgorithmsRealAlgebraic2006basu}. The $j$-th subresultant coefficient of $P$ and $P^{\prime}$, $\mathrm{sRes}_{j}(P,P^{\prime})$, is defined as follows:
\begin{equation}
    \mathrm{sRes}_{j}(P,P^{\prime})=D_{n-j},
\end{equation}
where $D_{n-j}$ follows the same definition in Eq~\eqref{eq:dlist}. Although $\mathrm{sRes}_{j}(P,P^{\prime})$ under this definition differs in sign from the definition commonly used, e.g. Ref.~\cite{AlgorithmsRealAlgebraic2006basu}, such a difference does not affect the application of the following theorem:
\begin{thm}
\label{subresultants}
Let $P$ and $Q$ be two non-zero polynomials of degree $p$ and $q$ and let $0\le j \le \operatorname{min}(p,q)$, then $\mathrm{deg}\left(\mathrm{gcd}(P(x),P^{\prime}(x))\right)\ge j$ if and only if
\begin{equation}
    \mathrm{sRes}_{0}(P,Q)=\cdots=\mathrm{sRes}_{j-1}(P,Q)=0.
\end{equation}
\end{thm}

The detailed proof of Theorem~\ref{subresultants} can be found in Ref.~\cite{AlgorithmsRealAlgebraic2006basu}. As a result of the above discussion and Theorem~\ref{subresultants}, the appearance of a CPDP requires two polynomial equations $\mathrm{sRes}_{0}(P,Q)$ and $\mathrm{sRes}_{1}(P,Q)$ to be satisfied. For a Hamiltonian with $\mathrm{PT}$ or $\mathrm{psH}$ symmetry, the characteristic polynomial is a real coefficient polynomial, making $\mathrm{sRes}_{0}(P,Q)$ and $\mathrm{sRes}_{1}(P,Q)$ two real equations. Hence the codimension of a CPDP is 2 in general.

\subsection{{IV} Homotopy details for the situation where \texorpdfstring{$\mu=2n$}{2} while \texorpdfstring{$m$}{2} varies from \texorpdfstring{$n$}{2} to \texorpdfstring{$n-1$}{2}}
The fundamental group of $\hat{X}^{(n-1)}_{0}=\text{UConf}_{n-1}(\mathbb{C}_{+}^{0})\times(\mathbb{R}^2/S_{2})$ can be obtained as follows (omitting the base point notation):
\begin{equation}
\begin{aligned}
    \pi_1(\hat{X}^{(n-1)}_{0}) &=\pi_1\left(\text{UConf}_{n-1}(\mathbb{C}_{+}^{0})\times(\mathbb{R}^2/S_{2})\right)\\
    &=\pi_1\left(\text{UConf}_{n-1}(\mathbb{C}_{+}^{0})\right)\times\pi_1\left(\mathbb{R}^2/S_{2}\right)\\
    &=\pi_1\left(\text{UConf}_{n-1}(\mathbb{C}_{+}^{0})\right)\times\pi_1\left(\mathbb{R}\times\mathbb{R}_{+}\right)\\
    &=\mathcal{B}_{n-1},
\end{aligned}
\end{equation}
where we use the facts that $\mathbb{R}^2/S_{2} \cong \mathbb{R}\times\mathbb{R}_{+}$ and $\mathbb{R}\times\mathbb{R}_{+}$ is simply connected, i.e., $\pi_1\left(\mathbb{R}\times\mathbb{R}_{+}\right)=0$.

The fundamental group of $X_{0}^{(n)}\cap \hat{X}_{0}^{(n-1)}$ (omitting the base point notation):
\begin{equation}
    \begin{aligned}
    \pi_1(X_{0}^{(n)}\cap \hat{X}_{0}^{(n-1)}) &=\pi_1\left((\text{Conf}_{n-1}\left(\mathbb{C}_{+}^{0}\right)/S_{n-1})\times \mathbb{R}\right)\\
    &=\pi_1\left(\text{UConf}_{n-1}(\mathbb{C}_{+}^{0})\right)\times\pi_1\left(\mathbb{R}\right)\\
    &=\mathcal{B}_{n-1}.
\end{aligned}
\end{equation}
\subsubsection*{The free product with amalgamation}
A \textbf{word} in $G$ and $H$ is a product of the form:
$s_1s_2\dots s_n$,
where $s_i$ is an element in $G$ or $H$. The \textbf{reduced words} is a word being reduced by the following rules:
\begin{itemize}
    \item Remove the identity element (of either G or H).
    \item Replace a pair $g_ig_j$ by its product in $G$, or a pair $h_ih_j$  by its product in $H$ where $g_i\in G$ and $h_i\in H$.
\end{itemize}
The \textbf{free product} of two groups $G$ and $H$ is a group whose elements are the reduced words in $G$ and $H$. 

\textbf{Amalgamation}: Consider two group $G$ and $H$ along with injective group homomorphisms:
\begin{equation}
    \varphi :F\rightarrow G \quad \text{and } \psi :F\rightarrow H, 
\end{equation}
where $F$ is a group. Then the free product with amalgamation $G*_F H$ is the free product of $G$ and $H$ adjoining as relations:
\begin{equation}
    \varphi(f)\psi(f)^{-1}=1, \quad \text{for all } f\in F.
\end{equation}
Thus, we have $\mathcal{B}_n*_{\mathcal{B}_{n-1}} \mathcal{B}_{n-1}=\mathcal{B}_n$.
\subsection{{V} Generalization to the \texorpdfstring{$\mu=2n+1$}{2} situation while \texorpdfstring{$m$}{2} varies from \texorpdfstring{$n$}{2} to \texorpdfstring{$n-1$}{2}}
For $\mu=2n+1$, we still consider the eigenvalue space where $m=n$ or $m=n-1$ denoted as $Y^{(n,n-1)}$ with the eigenvalue list:
\begin{equation}
  Y^{(n,n-1)}_0=\{[\epsilon_1,\ldots,\epsilon_{n-1},\epsilon_{n-1}^{\ast},\ldots,\epsilon_1^{\ast},\hat{\epsilon}_1,\hat{\epsilon}_2,\hat{\epsilon}_3]\}.
\end{equation}
Owing to the unorderness, we let $\hat{\epsilon}_2=\hat{\epsilon}_1^{\ast}$, $\text{Im}(\epsilon_i)\ge 0$, $\text{Im}(\hat{\epsilon}_1)\ge 0$, $\text{Im}(\hat{\epsilon}_3)= 0$ when $m=n$; and $\text{Im}(\hat{\epsilon_i})=0$, $\text{Re}(\hat{\epsilon}_i)\neq\text{Re}(\hat{\epsilon}_j)$ ($i,j\in\{1,2,3\}$ and $i\neq j$), $\text{Im}(\epsilon_i)> 0$ when $m=n-1$. Thus, we have $\hat{\epsilon}_1\in\mathbb{C}_{+}$ and $\hat{\epsilon}_2\in\mathbb{C}_{-}$. Since $\mathbb{C}_{+}\cong\mathbb{C}_{-}$ is equivalent to the closed half plane equipped half-disk topology, we define the following set in $\hat{Y}^{(n-1)}$:
\begin{equation}
\begin{alignedat}{3}
        \hat{Y}^{(n-1)}&= \{&&[\epsilon_1,\ldots,\epsilon_{n-1},\epsilon_{n-1}^{\ast},\ldots,\epsilon_1^{\ast},\tilde{\epsilon}_1,\tilde{\epsilon}_2,\tilde{\epsilon}_3]\\&\quad &&\mid \text{Im}(\epsilon_i)>0, \tilde{\epsilon}_i\in\mathbb{R}\}\\
        &=&&((\mathbb{C}_{+}^{0})^{n-1}/S_{n-1})\times (\mathbb{R}^3/S_{3}).
\end{alignedat}
\end{equation}
Now, the eigenvalue space $Y^{(n,n-1)}$ is the union of $X^{(n)}$ [defined in Eq.~(3) of the main text] and $\hat{Y}^{(n-1)}$. And $X^{(n)}\cap \hat{Y}^{(n-1)}=\{[\epsilon_1,\ldots,\epsilon_{n-1},\epsilon_{n-1}^{\ast},\ldots,\epsilon_1^{\ast},\tilde{\epsilon}_1,\tilde{\epsilon}_1,\tilde{\epsilon}_3] \mid \text{Im}(\epsilon_i)>0, \text{ }\tilde{\epsilon}_i\in\mathbb{R}\}=((\mathbb{C}_{+}^{0})^{n-1}/S_{n-1})\times \mathbb{R}^2$. Now, we remove CPDPs from the eigenvalue space (denoted with the subscript $0$), obtaining $X_{0}^{(n)}=\text{UConf}_{n}(\mathbb{C}_{+})\times\mathbb{R}$, $\hat{Y}^{(n-1)}_{0}=\text{UConf}_{n-1}(\mathbb{C}_{+}^{0})\times(\mathbb{R}^3/S_{3})$ and $Y^{(n,n-1)}_{0}=X_{0}^{(n)}\cup \hat{Y}_{0}^{(n-1)}$.  Since the above sets are open and path-connected (because of the half-disk topology), by Seifert–Van Kampen theorem~\cite{AlgebraicTopology2002hatcher}, taking $q_0\in X^{(n)}_{0}\cap \hat{Y}^{(n-1)}_{0}$ as the base point, the fundamental group of $Y^{(n,n-1)}_{0}$ is isomorphic to the free product of the fundamental group of $X_{0}^{(n)}$, $\hat{Y}^{(n-1)}_{0}$ with amalgamation of $\pi_{1}(X_{0}^{(n)}\cap \hat{Y}^{(n-1)}_{0},q_0)$:

\begin{equation}
\begin{aligned}
        &\pi_{1}(Y_{0}^{(n,n-1)},q_0)  \\&=\pi_{1}(X_{0}^{(n)},q_0)*_{\pi_{1}(X_{0}^{(n)}\cap\hat{Y}^{(n-1)}_{0},q_0)}\pi_{1}(\hat{Y}_{0}^{(n-1)},q_0) \\ &=\mathcal{B}_{n}.
\end{aligned}
\end{equation}

\subsection{{VI} Detailed math of the braid group }
\begin{figure}
\centering
	\includegraphics[width=1.0\columnwidth]{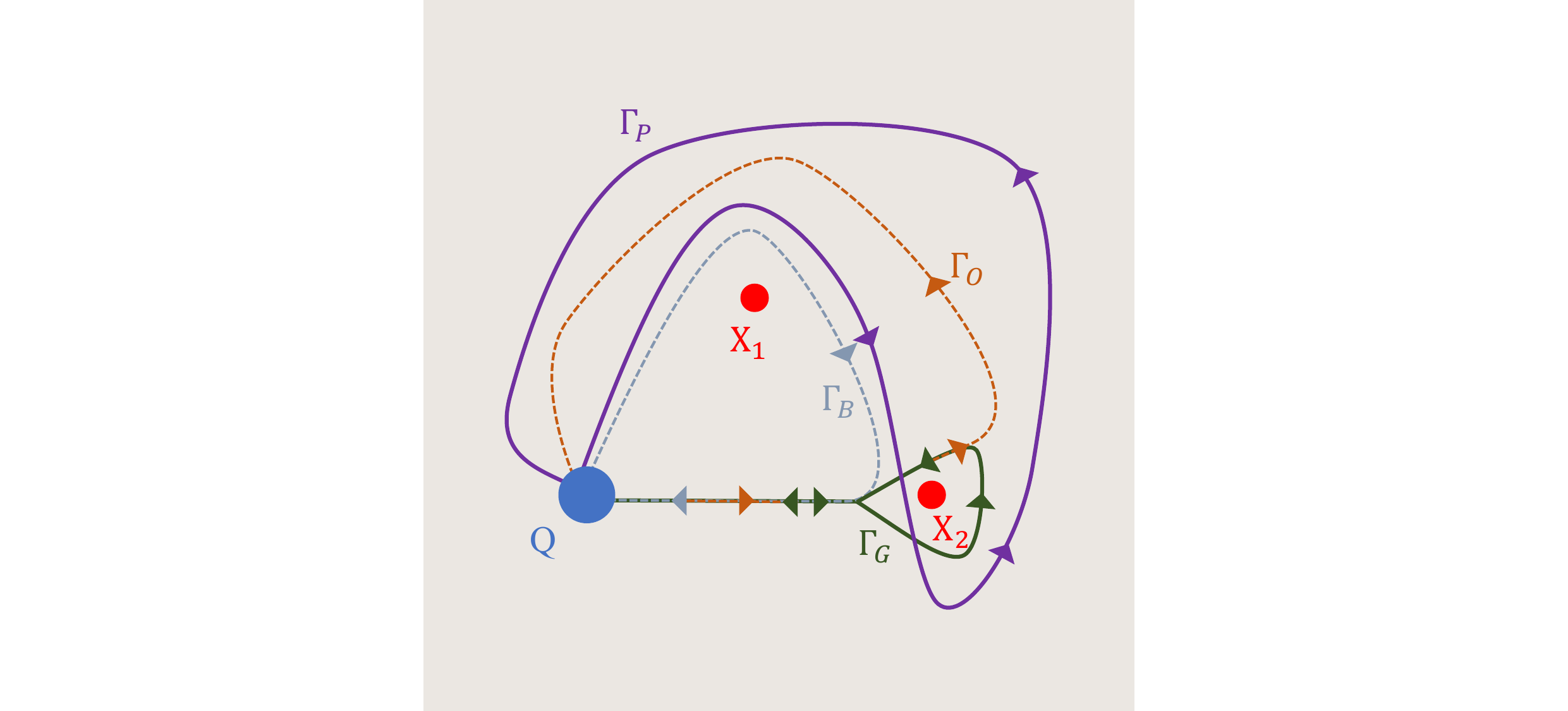}
	\caption {Paths (purple and green) originating from the same base point $Q$ and enclosing the same CPDP $\mathrm{X}_1$ could be not equivalent. Red Points are CPDPs.}
	\label{sup-fig1}
\end{figure}
The generators of Braid group $\mathcal{B}_n$ satisfy the braid relations~\cite{BraidGroups2008kassel}:
\begin{equation}
            \begin{cases}
            &\tau_i\tau_j=\tau_j\tau_i, \quad \text{if } |j-i|>1; \\
            &\tau_i\tau_{i+1}\tau_i=\tau_{i+1}\tau_{i}\tau_{i+1}, \quad \text{for all } 1\le i\le n-2.
            \end{cases}
\end{equation}
\subsubsection*{The conjugate relationship}
We consider a few paths as shown in Fig.~\ref{sup-fig1}. The green path $\Gamma_{G}$ may not be equivalent to the purple path $\Gamma_{P}$, though they share the same base point $Q$ and enclose the same CPDP $\mathrm{X}_2$. That is because the continuous transformation between $\Gamma_{G}$ and $\Gamma_{P}$ has to pass through another CPDP $\mathrm{X}_1$. Actually $\Gamma_{G}$ and $\Gamma_{P}$ have the following relationship:
\begin{equation}
     [\Gamma_P]=[\Gamma_{B}]\cdot [\Gamma_G]\cdot[\Gamma_O],
\end{equation}
where $\Gamma_{B}$, $\Gamma_{O}$ are the dashed blue-gray path and the dashed orange path in Fig.~\ref{sup-fig1}, respectively. With this equation, their braid words satisfy:
\begin{equation}
\begin{aligned}
            b_{\Gamma_P} &=b_{\Gamma_{B}} b_{\Gamma_G} b_{\Gamma_O} \\
    &=b_{\mathrm{X}_1}^{-1}b_{\Gamma_G}b_{\mathrm{X}_1},
\end{aligned}
\end{equation}
where $b_{i}$ denotes the braid invariant of a path $i$ or a CPDP $i$. Thus, the braid words of two paths follow a conjugate relationship. 

In the Fig.~2(c) of the main text, the blue CPDP line crosses above the red CPDP line. Hence, the continuous transformation between the two paths with base point $Q$ enclosing the red CPDP before and after the crossing (the blue CPDP line over the red CPDP line) has to pass through the blue CPDP. Thus, the braid word of the red CPDP after the crossing follows the conjugate relationship: 
\begin{equation}
\bar{b}_{red}=(\tau_1\tau_2^{-1}\tau_1^{-1})^{-1}\tau_1(\tau_1\tau_2^{-1}\tau_1^{-1})=\tau_2.
\end{equation}
Then correspondently, the braid word of the path $\Gamma$ at $t=t_1$ in Fig.~2(c) of the main text can be calculated as follows:
\begin{equation}
    \begin{aligned}
b_{\Gamma}(t_1)&=b_{dyn}^{-1}\bar{b}_{red}b_{green}b_{dyn}\\
&=(\tau_1\tau_2\tau_1\tau_2)^{-1}(\tau_1\tau_2^{-1}\tau_1^{-1})^{-1}\tau_1(\tau_1\tau_2^{-1}\tau_1^{-1})\tau_1^{-1}(\tau_1\tau_2\tau_1\tau_2)\\
&=\tau_2^{-1}\tau_1^{-1}\tau_2^{-1}\tau_1^{-1}\tau_1\tau_2\tau_1^{-1}\tau_1\tau_1\tau_2^{-1}\tau_1^{-1}\tau_1^{-1}\tau_1\tau_2\tau_1\tau_2\\
&=\tau_2^{-1}\tau_2^{-1}\tau_1^{-1}\tau_2^{-1}\tau_1\tau_2\tau_1\tau_2^{-1}\tau_1^{-1}\tau_2\tau_1\tau_2\\
&=\tau_2^{-1}\tau_2^{-1}\tau_1^{-1}\tau_2^{-1}\tau_2\tau_1\tau_2\tau_2^{-1}\tau_1^{-1}\tau_1\tau_2\tau_1\\
&=\tau_2^{-1}\tau_1.
    \end{aligned}
\end{equation}

\subsection{{VII} Electric circuits for the \texorpdfstring{$\mathrm{PT}^2=1$}{2} case}

\begin{figure}
\centering
	\includegraphics[width=1.0\columnwidth]{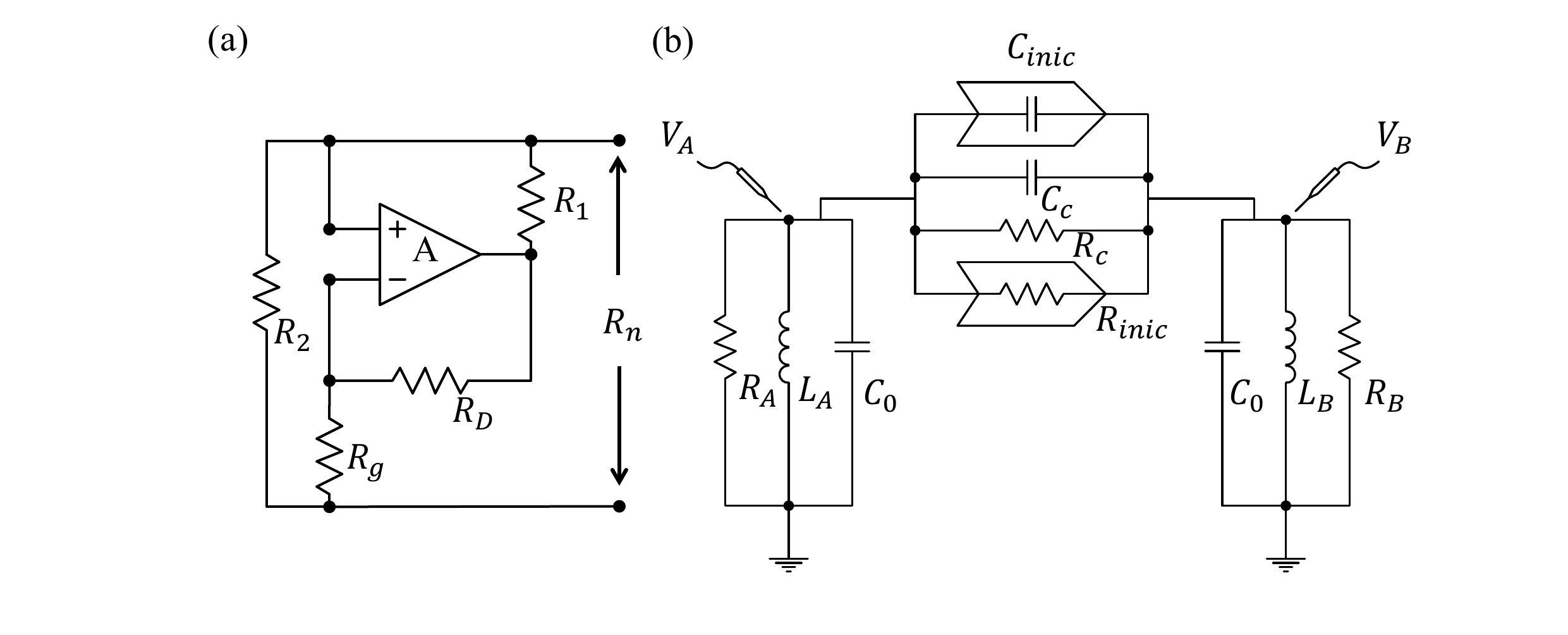}
	\caption {(a) Circuit used to realize an effective negative resistor. The negative resistance is realized with an amplifier feedback circuit~\cite{EnhancedSensingNondegraded2019xiao,NonlinearityenabledHigherorderExceptional2023bai}. (b) Circuit used to realize an arbitrary coupling between two LC resonance cavities. The coupling part consists of four independent tunable elements: a negative impedance converter with current inversion (INIC) associated with a capacitor $C_{\text{inic}}$~\cite{GeneralizedBulkBoundary2020helbig, CircuitsFiltersHandbook2009chen,NonHermitianSwallowtailCatastrophe2023hu}, a capacitor $C_c$, a resistor $R_c$, and an INIC associated with a resistor $R_{\text{inic}}$~\cite{ChiralVoltagePropagation2019hofmann,CircuitsFiltersHandbook2009chen}.}
	\label{sup-fig2}
\end{figure}
In this section, we demonstrate that the circuit system can be implemented to realize our model presented in the main text. Firstly, an effective negative resistor can be realized with the circuit in Fig.~\ref{sup-fig2}(a) with a negative resistance~\cite{EnhancedSensingNondegraded2019xiao,NonlinearityenabledHigherorderExceptional2023bai}
\begin{equation}
R_n=-(R_2R_gR_1/R_D)/(R_2-R_gR_1/R_D). 
\end{equation}
In addition, four circuit elements together are capable to achieve an arbitrary coupling as shown in Fig.~\ref{sup-fig2}(b). Kirchoff's Equation for the circuit in Fig.~\ref{sup-fig2}(b) are:
\begin{equation}
    \begin{aligned}
& \frac{V_A}{-i \omega L_A}+\frac{V_A}{R_A}-i \omega C_0 V_A-i \omega\left(C_c-C_{\text {inic }}\right)\left(V_A-V_B\right)+\frac{V_A-V_B}{R_c}-\frac{V_A-V_B}{R_{\text {inic }}}=0, \\
& \frac{V_B}{-i \omega L_B}+\frac{V_B}{R_B}-i \omega C_0 V_B-i \omega\left(C_c+C_{\text {inic }}\right)\left(V_B-V_A\right)+\frac{V_B-V_A}{R_c}+\frac{V_B-V_A}{R_{\text {inic }}}=0,
\end{aligned}
\end{equation}
where $R_{A,B}$ are the effective resistances such as that from the negative resistor shown in Fig.~\ref{sup-fig2}(a). We denote the resonance frequencies of the uncoupled LC circuit $\omega_A=1/\sqrt{L_AC_0}$ and reformulate the equations in a matrix form as:
\begin{equation}
\label{eq:matrixkirch}
\left(\begin{array}{cc}
Y_A & \frac{i \omega}{2 C_0}\left(\frac{1}{R_c}-\frac{1}{R_{\text {inic }}}\right)+\omega^2 \frac{C_c-C_{\text {inic }}}{2 C_0} \\
\frac{i \omega}{2 C_0}\left(\frac{1}{R_c}+\frac{1}{R_{\text {inic }}}\right)+\omega^2 \frac{C_c+C_{\text {inic }}}{2 C_0} & Y_B
\end{array}\right)\left(\begin{array}{l}
V_A \\
V_B
\end{array}\right)=0,
\end{equation}
where $Y_A=-\frac{i \omega}{2 C_0 R_A}+\frac{\omega_A^2-\omega^2}{2}-\omega^2 \frac{C_C-C_{\text {inic }}}{2 C_0}-\frac{i \omega}{2 C_0}\left(\frac{1}{R_C}-\frac{1}{R_{\text {inic }}}\right)$ and $Y_B=-\frac{i \omega}{2 C_0 R_B}+\frac{\omega_B^2-\omega^2}{2}-\omega^2 \frac{C_C+C_{\text {inic }}}{2 C_0}-\frac{i \omega}{2 C_0}\left(\frac{1}{R_C}+\frac{1}{R_{\text {inic }}}\right)$. Take the approximations that $\{C_c,C_{\text{inic}}\}\ll C_0$, $\{R_A,R_B\}\ll {R_C,R_{\text{inic}}}$ and $|\omega_{A,B}-\omega|\ll \omega$, then Eq.~\eqref{eq:matrixkirch} becomes:
\begin{equation}
\label{eq:circuithami}
\left(\begin{array}{cc}
\omega_A-\frac{i}{2 R_A C_0} & \frac{i}{2 C_0}\left(\frac{1}{R_c}-\frac{1}{R_{\text {inic }}}\right)+\omega_B \frac{C_c-C_{\text {inic }}}{2 C_0} \\
\frac{i}{2 C_0}\left(\frac{1}{R_c}+\frac{1}{R_{\text {inic }}}\right)+\omega_B \frac{C_c+C_{\text {inic }}}{2 C_0} & \omega_B-\frac{i}{2 C_0 R_B}
\end{array}\right)\left(\begin{array}{l}
V_A \\
V_B
\end{array}\right)=\omega\left(\begin{array}{l}
V_A \\
V_B
\end{array}\right),
\end{equation}
which is consistent with Eq.~(16) in the main text. Thus, with the four circuit elements, one can realize an arbitrary coupling as required in the Hamiltonian Eq.~(14) of the main text. 

\subsubsection*{The parameter setting of the circuit system}

\begin{figure}
\centering
	\includegraphics[width=1.0\columnwidth]{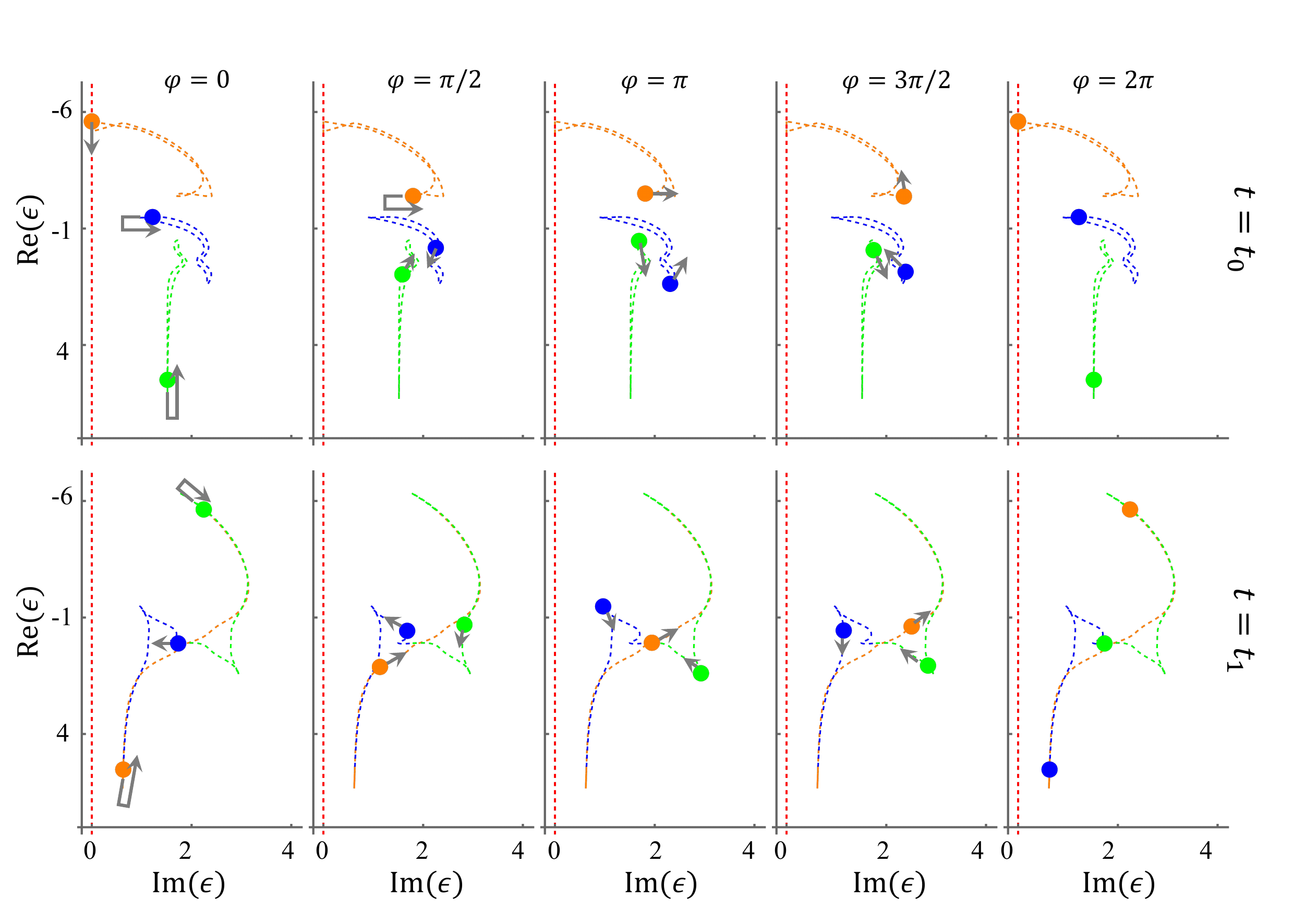}
	\caption {Configurations of the first $n$ eigenvalues on the complex plane with their past trajectory at a few $\phi$ instances for $t=t_0$ and $t=t_1$. Here we parameterize the circular path $\Gamma$ by $(R_{2,A},L_{1,A})=(1/4+\cos(\phi+7/5)/19,1/16+\sin(\phi+7/5)/19)$. These configurations correspond to the Fig.~2(e) in the main text. The red dashed lines denote the real axis. All the parameters are dimensionless.}
	\label{sup-fig3}
\end{figure}

We set $C_i=C_0=1$ as normalized parameter and introduce $\tilde{\kappa}$ with the frequency unit. Thus, capacitance, inductance, resistance, frequency are nondimensionalized by $C_0$, $1/(\tilde{\kappa}^2C_0)$, $1/(\tilde{\kappa}C_0)$, $\tilde{\kappa}$ respectively and become dimensionless.  For the circuit elements,  we set $L_{i,A}=L_{i,B}$, $\omega_i=\omega_{i,B}=\omega_{i,A}=1/\sqrt{L_{i,A}C_0}$, $R_{i,B}=-R_{i,A}$, $R_{1,\text{inic}}=R_{1,c}=+\infty$, $C_{1,c}=C_{1,\text{inic}}=0$, $R_{2,c}=R_{3,c}=+\infty$, $C_{2,\text{inic}}=C_{3,\text{inic}}=0$ where $i\in\{1,2,3\}$. Now, we use the same scheme to realize coupling between boxes [$H_i$ in Eq.~(14) of the main text]. We use the subscript $13$ ($32$) and $AB$ or $BA$ to represent the coupling from the cavity $A$ or $B$ of $H_{1}$ ($H_{3}$) to the cavity $B$ or $A$ of $H_{3}$ ($H_{2}$). $C^{0}$ is set to realize the identical coupling marked by the double black arrow in Fig.~2(a) of the main text. The parameters appeared in the circuit equation Eq.~(16) of the main text are set as follows: $R_{j,k,\text{inic}}=R_{j,k,c}$, $C_{j,k,\text{inic}}=C_{j,k,c}$, $C_{j,BA,c}=C_{j,AB,c}$, $R_{i,BA,c}=-R_{i,AB,c}$ where $j\in\{13,32\}$, $k\in\{AB,BA\}$. So, the Hamiltonian of the model [Eq.~(13) in the main text] can be realized in the circuit system presented in Fig.~2(b) of the main text. 

To observe path-dependent annihilation of CPDPs, we set $C_{2,c}=-\sqrt{L_{2,A}}(1-4 R_{2,A})/(2R_{2,A})$, $R_{2,\text{inic}}=4R_{2,A}/(1-4R_{2,A})$, $R_{3,A}=R_{1,A}/(8R_{1,A}-1)$, $L_{3,A}=1/(8-1/\sqrt{L_{1,A}})^2$, $C_{3,c}=-3(4\sqrt{L_{1,A}}-1)/(8\sqrt{L_{1,A}}-1)/2/\sqrt{2}$, $R_{3,\text{inic}}=2\sqrt{2}/(4-1/\sqrt{L_{1,A}})$, $C_{c}^{13,AB}=\sqrt{L_{1,A}}(4R_{2,A}-1)/8/R_{2,A}$, $R_{13,AB,c}=8R_{2,A}/(3-12R_{2,A})$, $C_{32,AB,c}=-5(4\sqrt{L_{1,A}}-1)/7\sqrt{2}/(8\sqrt{L_{1,A}}-1)$, $R_{32,AB,c}=7\sqrt{2}/5/(4-1/\sqrt{L_{1,A}})$. Thus all the parameters are functions of $R_{1,A}$ and $L_{2,A}$. Then we let $R_{1,A}$ and $L_{2,A}$ vary as a function of time: $R_{1,A}=1/(4+2\sqrt{2}-5t/\sqrt{2})$, $L_{2,A}=1/(-44+70t-25t^2)^2$. $t_0= 1.08$, $t_1 = 1.59$ in Fig.~2(c) of the main text. The base point $Q_0$ is on the nodal line and at $(R_{2,A},L_{1,A}) =(6/25,3/250)$. All capacitances, inductances, resistances, frequencies are nondimensionalized by $C_0$, $1/(\tilde{\kappa}^2C_0)$, $1/(\tilde{\kappa}C_0)$, $\tilde{\kappa}$ respectively and become dimensionless.

\begin{figure}
	\centering
	\includegraphics[width=1.0\columnwidth]{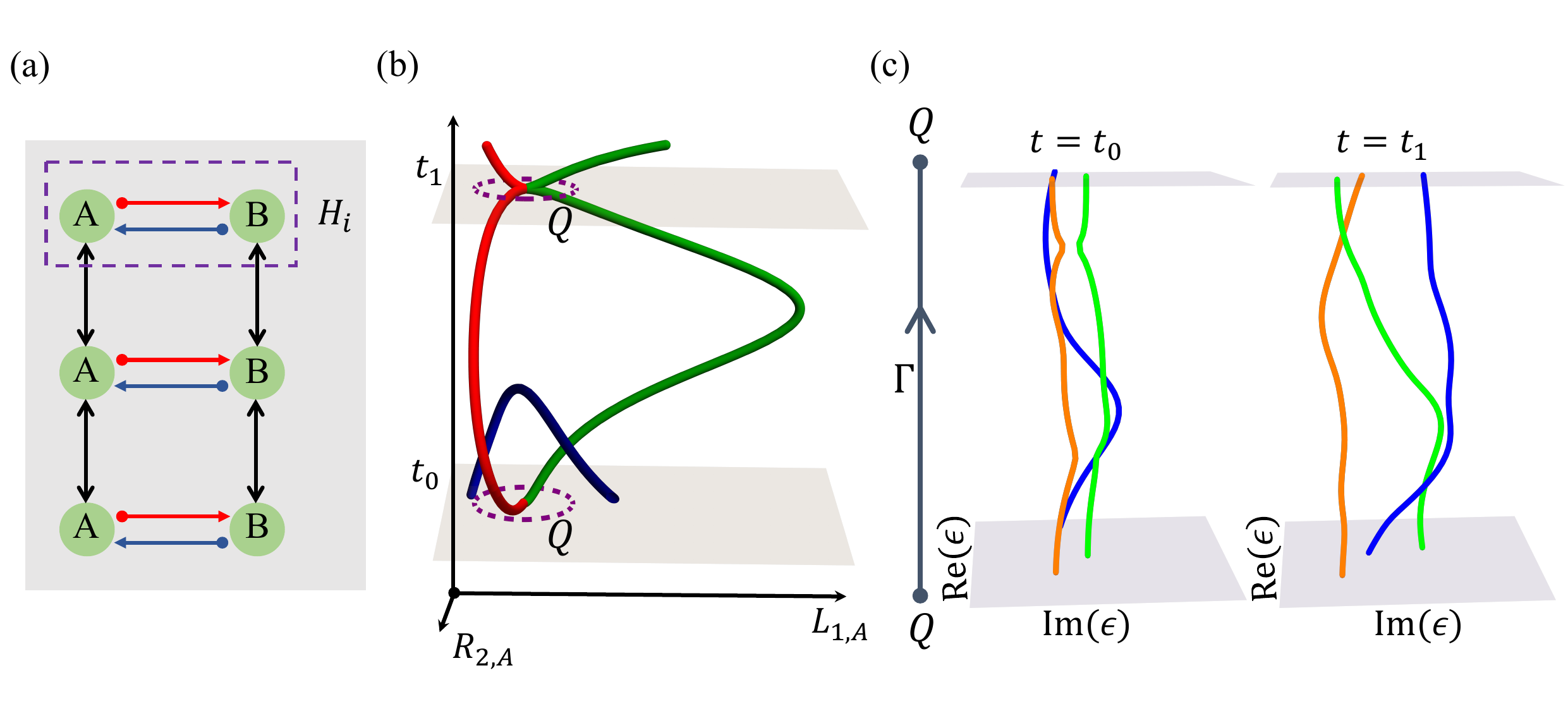}\\
	\caption {(a) An electric circuit system realizes our model. The circuit subsystem enclosed by the purple dashed box is equivalent to the $H_i$ in Eq.~\eqref{eq:ptn1hi}. (b) Path-dependent annihilation of CPDP lines in the $(R_{2,A},L_{1,A},t)$ space. Different lines are distinguished by colors. The purple dashed line denotes the path $\Gamma$ originating from point $Q$. We parameterize $\Gamma$ by $(R_{2,A},L_{1,A})=(1/4+\cos(\phi)/10,4/25+\sin(\phi)/10)$ (c) The first $3$ eigenvalue strands braiding along the path $\Gamma$ in (b), at $t=t_0$ and $t=t_1$. The base point $Q$ is at $(R_{2,A},L_{1,A})= (3/10,1/5)$. The green, blue, and orange lines represent eigenvalues with decreasing imaginary parts at $Q$. Their projective trajectories are presented in Fig.~\ref{sup-fig5}.
}
	\label{sup-fig4}
\end{figure}

\subsection{{VIII} The \texorpdfstring{$\mathrm{PT}^2=-1$}{2} case and its corresponding circuit system}
In this section, we present a model with $\mathrm{PT}^2=-1$ and its circuit simulation.
Considering a six-state model with the following Hamiltonian:
\begin{equation}
  \label{eq:ptn1}
  H = \left(\begin{array}{ccc}
    H_1 & \mathbb{I}_2 & 0 \\
    \mathbb{I}_2 & H_2 & \mathbb{I}_2 \\
    0 & \mathbb{I}_2 & H_3
  \end{array}\right),
\end{equation}
where,
\begin{equation}
  \label{eq:ptn1hi}
  H_i = \left(\begin{array}{cc}
    \omega_i+i l_i & \kappa_i\\
    -\kappa_i^{\ast} & \omega_i-i l_i
  \end{array}\right), 
\end{equation}
and $\omega_i,l_i \in \mathbb{R}$, $\kappa \in \mathbb{C}$. $H$ satisfies $\mathrm{PT}$ symmetry, with:
\begin{equation}
  U_{\mathrm{PT}}=\mathbb{I}_3\otimes \sigma _y,\quad U_{\mathrm{PT}}U_{\mathrm{PT}}^{\ast}=-1.
\end{equation}
This Hamiltonian $H$ can be realized by coupling three two-level subsystems $H_i$. $H_i$ can be realized by the circuit subsystem outlined by purple dashed boxes in Fig.~\ref{sup-fig4}(a). Comparing the circuit equation Eq.~\eqref{eq:circuithami} with the Hamiltonian $H_i$ in Eq.~\eqref{eq:ptn1hi}, we can set $L_{i,A}=L_{i,B}$, $\omega_{i}=\omega_{i,B}=\omega_{i,A}=1/\sqrt{L_{i,A}C_0}$, $C_{i,c}=0$, $R_{i,B}=-R_{i,A}$, $R_{i,\text{inic}}=+\infty$, $\kappa_i=\frac{i }{2 C_0 R_{i,c}}-\omega_{i,B} \frac{C_{i,\text{inic}}}{2 C_0}$ and $l_i=\frac{i}{2 R_{i,B} C_0}$. So, the Hamiltonian in Eq.~\eqref{eq:ptn1} can be realized in the circuit system presented in Fig.~\ref{sup-fig4}(a). 

\begin{figure}
	\centering
	\includegraphics[width=1.0\columnwidth]{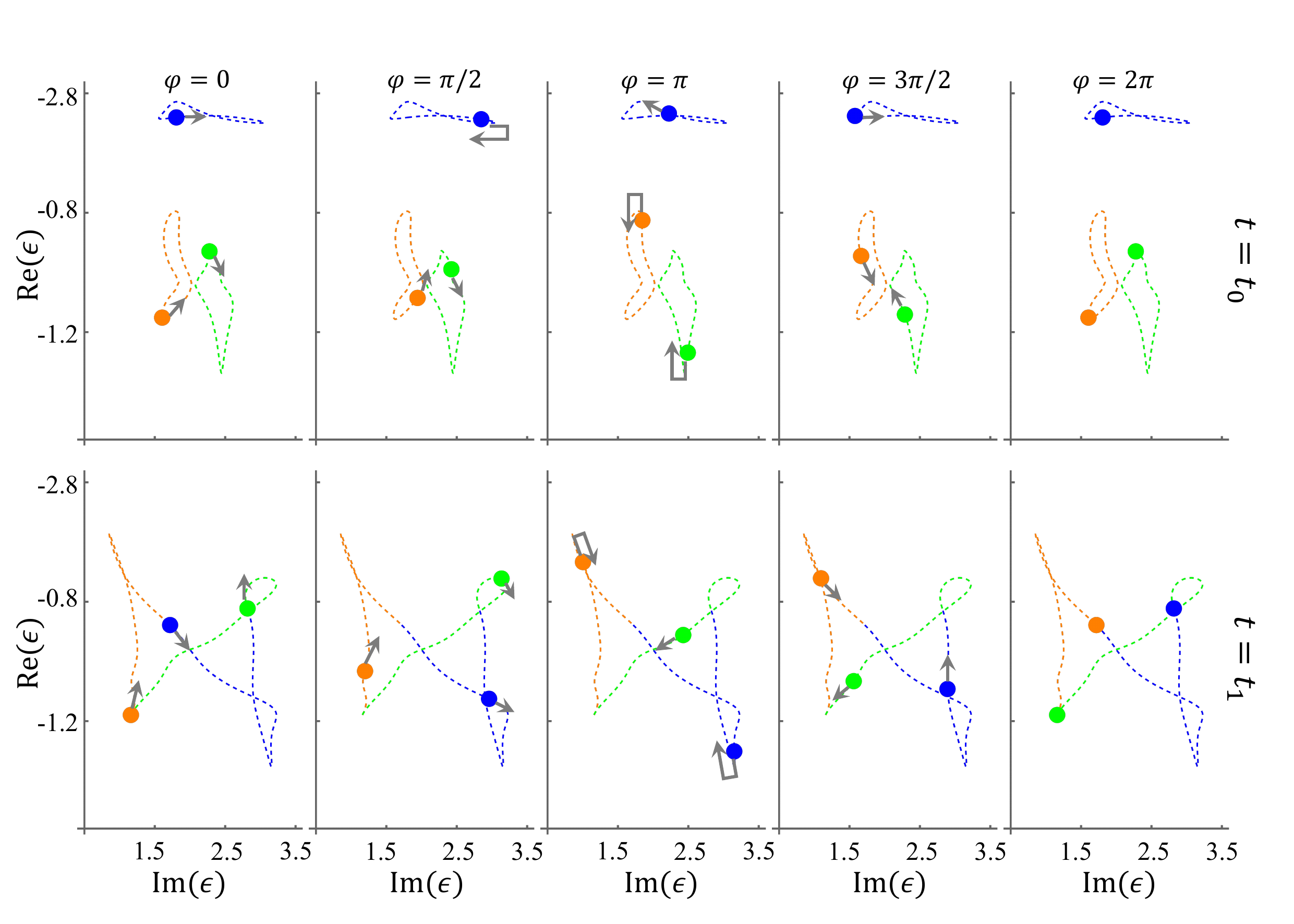}\\
	\caption {Configurations of the first $n$ eigenvalues over the complex plane with their past trajectory at a few $\phi$ instances for $t=t_0$ and $t=t_1$, where we parameterize the path $\Gamma$ by $(R_{2,A},L_{1,A})=(1/4+\cos(\phi)/10,4/25+\sin(\phi)/10)$. These configurations correspond to the Fig.~\ref{sup-fig4}(c). }
	\label{sup-fig5}
\end{figure}

To observe path-dependent annihilation of CPDPs, the parameters are set as: $C_{1,\text{inic}}=\sqrt{L_{1,A}}(1-4R_{2,A})/2R_{2,A}$, $C_{\text{inic}}^{2}=0$, $R_{c}^{2}=2\sqrt{2L_{A}^{1}}/(-2+5L_{A}^{1})$, $C_{c}^{3}=0$, $R_{c}^{3}=+\infty$, $C_{\text{inic}}^{3}=5\sqrt{L_{3,A}}(1-4R_{2,A})/9R_{2,A}$, $R_{3,A}=R_{1,A}/(8R_{1,A}-1)$. $R_{1,A}$ and $L_{2,A}$ vary as a function of $t$: $R_{1,A}=2/(-3+4\sqrt{2}+4t)$, $L_{2,A}=4/(65+32(-3+t)t)^2$. All capacitances, inductances, resistances, frequencies are nondimensionalized by $C_0$, $1/(\tilde{\kappa}C_0)$, $1/(\tilde{\kappa}^2C_0)$,$\tilde{\kappa}$ respectively and become dimensionless.

We investigate the evolution of CPDPs in the 3D $(R_{2,A},L_{1,A},t)$ parameter space. A pair of CPDPs appears at $t_0 \approx 1.1$.  We associate the green one and the red one with the opposite braid words $\tau_1\tau_2\tau_1^{-1}$ and $\tau_1\tau_2^{-1}\tau_1^{-1}$. They subsequently follow a detour around the blue one along $t$ and merge at $t_1 = 1.75$, after which they split for $t>t_1$. When the blue nodal line with the braid word $\tau_1$ passes above the green nodal line, the braid words of the green CPDP changes to $\tau_2$, leading the braid invariant of the path $\Gamma$ at $t=t_1$ to be $\tau_1^{-1}\tau_2$. This results in the non-annihilation of the red and green nodal line when they merge at $t=t_1$. And their braid invariants at $t_0$ and $t_1$ are $1$ and $\tau_1^{-1}\tau_2$, respectively, as shown in Fig.~\ref{sup-fig4} and Fig.~\ref{sup-fig5}.

\end{document}